\documentclass[fleqn,usenatbib]{mnras}

\usepackage{mathptmx}

\usepackage[T1]{fontenc}
\usepackage{ae,aecompl}

\usepackage{graphics,graphicx}	
\usepackage{amsmath}	
\usepackage{amssymb}	
\usepackage{multirow}
\usepackage{placeins}
\usepackage{caption}
\usepackage{morefloats}
\usepackage{newtxtext,newtxmath}

\newcommand{\emm}[1]{\ensuremath{#1}}   
\newcommand{\emr}[1]{\emm{\mathrm{#1}}} 
\newcommand{\unit}[1]{\emm{\, \emr{#1}}}

\newcommand{\twCO}{\emr{^{12}CO}}
\newcommand{\thCO}{\emr{^{13}CO}}
\newcommand{\CeiO}{\emr{C^{18}O}}
\newcommand{\HCOp}{\emr{HCO^{+}}}
\newcommand{\Jone}{\mbox{(1--0)}}
\newcommand{\Jtwo}{\mbox{(2--1)}}

\newcommand{\Jy}{\unit{Jy}}
\newcommand{\K}{\unit{K}}
\newcommand{\mK}{\unit{mK}}

\newcommand{\kms}   {\unit{km\,s^{-1}}}
\newcommand{\MHz} {\unit{MHz}}
\newcommand{\GHz} {\unit{GHz}}
\newcommand{\kHz} {\unit{kHz}}
\newcommand{\pc}    {\unit{\pc}}

\renewcommand{\deg}{\emm{^\circ}}

\mathchardef\mhyphen="2D

\newcommand{\TabObsNOEMA}{%
  \begin{table}
  \centering{}
    \caption{NOEMA observation parameters.}
    \label{tab:obs}
    \centering{} 
    \resizebox{\columnwidth}{!}{
      \begin{tabular}{lcccc}
        \hline
        \hline
        Line & Beam & Beam & PA  & Noise$\mathrm{^{(a)}}$ \\
                 & [\arcsec] & [pc] & [\deg]  & [mK] \\
        \hline
         HCN\Jone & $1.87\times1.04$ & $102.4\times57.0$ & 26 & 51 \\
         HCO$^+$\Jone & $1.86\times1.03$ & $101.8\times56.4$ & 27  54 \\
         HNC\Jone  & $1.85\times1.05$ & $101.2\times57.5$ & 22 & 54 \\
        \CeiO\Jone & $1.52\times0.84$ & $83.2\times46$ & 27  & 78 \\
        \thCO\Jone & $1.52\times0.84$ & $83.2\times46$ & 27  & 87 \\
        \hline
      \end{tabular} }  
      \begin{minipage}{0.95\columnwidth}
        \vspace{1mm}
        (a) Evaluated at the mosaic phase centre, which is close to the galaxy centre (the noise steeply increases at the mosaic edges after correction for primary beam attenuation), at a spectral resolution of $5$\,\kms.
      \end{minipage}
  \end{table} 
  }

\newcommand{\GalaxyProp} {\begin{table}
\centering
\caption{Source information.}
\begin{tabular}{cc}
    \hline \hline
    Property & Value \\
    \hline
    Name & NGC~3627 (Messier 66) \\
    Hubble type $^{(a)}$ & SABb \\
    Centre RA (J2000) & 11h20m14.867s \\
    Centre DEC (J2000) & 12d59m34.05s  \\
    Inclination, $i$ $\mathrm{[^{\circ}]}$ $^{(a)}$ & 62 \\ 
    Position angle, PA $\mathrm{[^{\circ}]}$ $^{(a)}$ & 173 \\ 
    Distance, $D$ $\mathrm{[Mpc]}$ $^{(b)}$ & 11.3 \\
    r$_{25}$ [\arcmin] $^{(b,c)}$ & 5.1 \\
    $V_\mathrm{sys, hel}$ [km\,s$^{-1}$] $^{(d)}$ & 744 \\
    Metallicity [12+log(O/H)] $^{(e)}$ & $8.328\pm0.004$ \\
    $\left \langle \Sigma_\mathrm{SFR} \right \rangle$ [$\mathrm{M}_\odot$\,yr$^{-1}$\,kpc$^{-2}$] $^{(f)}$ & $7.7\times10^{-3}$ \\
    log$_{10}(M_*$) [$\mathrm{M}_\odot$] $^{(g)}$ & 10.5\\
    \hline\hline
\end{tabular}
\begin{minipage}{0.95\columnwidth}
    \vspace{1mm}
    (a) Morphology taken from the NASA Extragalactic
Database (NED) \\
    (b) Distance adopted from \citet{anand_2020}\\
    (c) Radius of the $B$-band 25th magnitude isophote\\
    (d) Systemic velocity from \citet{casasola_2011} \\
    (e) Metallicity calibration taken from \citet{pilyugin_thuan_2005} and metallicity value from \citet{kreckel_2019} \\
    (f) Average star formation rate surface density inside 0.75\,r$_{25}$, taken from the PHANGS-ALMA survey paper (Leroy et al, 2020, in prep)\\
    (g) Integrated stellar mass based on 3.6\,\micron\ emission, taken from the PHANGS-ALMA survey paper (Leroy et al, 2020, in prep)\\
\end{minipage}
\label{tab:NGC3627_prop}
\end{table} }

\newcommand{\Ncrit} { \begin{table}
\centering
 \caption{Properties of the observed molecular lines. We tabulate the molecular transition, rest frequency, and the effective critical density \citep{leroy_2017a} taken at a temperature of 25~K for lines studied here.}
\begin{tabular}{ lcc }
\hline \hline 
Line & $\nu_\mathrm{rest}$ $\mathrm{[GHz]}$ & $n_\mathrm{eff, crit}$ {[}$\mathrm{cm^{-3}}${]} \\
\hline
$^{12}$CO$\rm{^{(a, b)}}$ \Jtwo{} & 230.53 & $\mathrm{1\times10^3}$ \\
\thCO$\rm{^{(a, c)}}$ \Jone  & 110.20 & $\mathrm{8\times10^2}$ \\
\CeiO$\rm{^{(a, c)}}$ \Jone  & 109.78 & $\mathrm{8\times10^2}$ \\
\HCOp$\rm{^{(a, d)}}$ \Jone & 89.19 & $\mathrm{4\times10^4}$ \\
HCN$\rm{^{(a, d)}}$ \Jone & 88.63 & $\mathrm{2\times10^5}$  \\
HNC$\rm{^{(a, d)}}$ \Jone & 90.66  & $\mathrm{1\times10^5}$ \\
\hline \hline
\end{tabular}
\begin{minipage}{0.95\columnwidth}
    \vspace{1mm}
    (a) Calculated from the Leiden Atomic and Molecular Database (LAMDA) \citep{schoier_2005, vandertak_2007}. \\
    (b) The opacity of 100 adopted for this line (assuming the CO\Jone{} is optically thick \citep{leroy_2017a}).\\
    (c) We assume a fixed optical depth of 0.1\\
    (d) Here we assume a fixed optical depth of 1.
\end{minipage}
\label{tab:line_prop}
\end{table}
}

\newcommand{\UBonn}{Argelander-Institut f\"ur Astronomie, Universit\"at Bonn, Auf dem H\"ugel 71, 53121 Bonn, Germany}

\newcommand{\IRAM}{IRAM, 300 rue de la Piscine, 38406 Saint Martin d'H\`eres, France}

\newcommand{\LERMA}{LERMA, Observatoire de Paris, PSL Research University, CNRS, Sorbonne Universit\'es, 75014 Paris, France}

\newcommand{\MPI}{Max Planck Institut f{\"u}r Astronomie, K{\"o}nigstuhl 17, 69117 Heidelberg, Germany}

\newcommand{\INAF}{INAF -- Osservatorio Astrofisico di Arcetri, Largo E. Fermi 5, I-50157, Firenze, Italy}

\newcommand{\ARI}{Astronomisches Rechen-Institut, Zentrum f\"{u}r Astronomie der Universit\"{a}t Heidelberg, M\"{o}nchhofstra\ss e 12-14, 69120 Heidelberg, Germany}

\newcommand{\ObsMadrid}{Observatorio Astronómico Nacional (IGN), C/ Alfonso XII 3, E-28014 Madrid, Spain}

\newcommand{\UOhio}{Department of Astronomy, The Ohio State University, 4055 McPherson Laboratory, 140 West 18th Avenue, Columbus, OH 43210, USA}

\newcommand{\UAlberta}{4-183 CCIS, University of Alberta, Edmonton AB T6G 2E1, Alberta, Canada}

\newcommand{\UTolouse}{Université de Toulouse, UPS-OMP, F-31028 Toulouse, France ; CNRS, IRAP, Av. du Colonel Roche BP 44346, F-31028 Toulouse cedex 4, France}

\newcommand{\MPE}{Max Planck Institut f\"ur Extraterrestrische Physik, Giessenbachstra{\ss}e 1, D-85748 Garching, Germany}

\newcommand{\UGent}{Sterrenkundig Observatorium, Universiteit Gent, Krijgslaan 281 S9, B-9000 Gent, Belgium}

\newcommand{\ESO}{European Southern Observatory, Karl-Schwarzschild Stra{\ss}e 2, D-85748 Garching bei M\"{u}nchen, Germany}

\newcommand{\ITA}{Universit\"{a}t Heidelberg, Zentrum f\"{u}r Astronomie, Institut f\"{u}r Theoretische Astrophysik, Albert-Ueberle-Str 2, D-69120 Heidelberg, Germany}

\newcommand{\UMassachusets}{Department of Astronomy, University of Massachusetts Amherst, 710 North Pleasant Street, Amherst, MA 01003, USA}

\newcommand{\ULion}{Univ Lyon, Univ Lyon1, ENS de Lyon, CNRS, Centre de Recherche Astrophysique de Lyon UMR5574, F-69230 Saint-Genis-Laval France}

\newcommand{\UCanberra}{Research School of Astronomy and Astrophysics, Australian National University, Canberra, ACT 2611, Australia}

\newcommand{\ArcAus}{ARC Centre of Excellence for All Sky Astrophysics in 3 Dimensions (ASTRO 3D), Australia}

\newcommand{\IWR}{Universit\"{a}t Heidelberg, Interdisziplin\"{a}res Zentrum f\"{u}r Wissenschaftliches Rechnen, Im Neuenheimer Feld 205, D-69120 Heidelberg, Germany}

\title[NGC~3627 : dense gas tracers on molecular cloud scale]{Dense molecular gas properties on 100\,pc scales across the disc of NGC~3627}

\author[I.~Be\v{s}li\'c et al.]{I.~Be\v{s}li\'c$^{1}$,\thanks{E-mail: ibeslic@uni-bonn.de (IB)}
A.~T.~Barnes$^{1}$, F.~Bigiel$^{1}$, J.~Puschnig$^{1}$, J.~Pety$^{2,3}$, C.~Herrera Contreras$^{2}$,
\newauthor A.~K.~Leroy$^{4}$, A.~Usero$^{5}$, E.~Schinnerer$^{6}$, S.~E.~Meidt$^{7}$, E.~Emsellem$^{8,9}$, A.~Hughes$^{10}$, 
\newauthor C.~Faesi$^{6, 11}$, 
K.~Kreckel$^{12}$, F.~M.~C.~Belfiore$^{13}$, M.~Chevance$^{12}$, J.~S.~den Brok$^{1}$, 
\newauthor  C.~Eibensteiner$^{1}$, S.~C.~O.~Glover$^{14}$, K.~Grasha$^{15,16}$ M.~J.~Jimenez-Donaire$^{5}$, R.~S.~Klessen$^{14,17}$, 
\newauthor J.~M.~D.~Kruijssen$^{12}$, 
 D.~Liu$^{6}$, I.~Pessa$^{6}$, M.~Querejeta$^{5}$, E.~Rosolowsky$^{18}$, T.~Saito$^{6}$, 
  \newauthor  F.~Santoro$^{6}$, A.~Schruba$^{19}$, M.~C.~Sormani$^{14}$, and T.~G.~Williams$^{6}$
\\ \\
$^{1}$\UBonn \\
$^{2}$\IRAM \\
$^{3}$\LERMA \\
$^{4}$\UOhio \\
$^{5}$\ObsMadrid \\
$^{6}$\MPI \\
$^{7}$\UGent \\
$^{8}$\ESO \\
$^{9}$\ULion \\
$^{10}$\UTolouse \\
$^{11}$\UMassachusets \\
$^{12}$\ARI \\
$^{13}$\INAF \\
$^{14}$\ITA \\
$^{15}$\UCanberra \\
$^{16}$\ArcAus \\
$^{17}$\IWR\\
$^{18}$\UAlberta \\
$^{19}$\MPE
}

\date{Accepted XXX. Received YYY; in original form ZZZ}

\pubyear{2021}

\begin{document}
\label{firstpage}
\pagerange{\pageref{firstpage}--\pageref{lastpage}}
\maketitle

\begin{abstract}
It is still poorly constrained how the densest phase of the interstellar medium varies across galactic environment. A large observing time is required to recover significant emission from dense molecular gas at high spatial resolution, and to cover a large dynamic range of extragalactic disc environments. We present new NOrthern Extended Millimeter Array (NOEMA) observations of a range of high critical density molecular tracers (HCN, HNC, HCO$^{+}$) and CO isotopologues (\thCO{}, C$^{18}$O) towards the nearby ($11.3$~Mpc), strongly barred galaxy NGC~3627. These observations represent the current highest angular resolution (1.85\arcsec; 100~pc) map of dense gas tracers across a disc of a nearby spiral galaxy, which we use here to assess the properties of the dense molecular gas, and their variation as a function of galactocentric radius, molecular gas, and star formation. We find that the HCN\Jone/CO\Jtwo{}\ integrated intensity ratio does not correlate with the amount of recent star formation. Instead, the HCN\Jone/CO\Jtwo{}\ ratio depends on the galactic environment, with differences between the galaxy centre, bar, and bar end regions. The dense gas in the central 600~pc appears to produce stars less efficiently despite containing a higher fraction of dense molecular gas than the bar ends where the star formation is enhanced. In assessing the dynamics of the dense gas, we find the HCN\Jone\ and \HCOp\Jone\ emission lines showing multiple components towards regions in the bar ends that correspond to previously identified features in CO emission. These features are co-spatial with peaks of H$\alpha$ emission, which highlights that the complex dynamics of this bar end region could be linked to local enhancements in the star formation.
\end{abstract}



\section{Introduction}

Star formation occurs in the coldest, densest parts of molecular clouds. This is observed within star-forming regions in the Milky Way, where it has been shown that the star formation rate (SFR) surface density ($\mathrm{\Sigma_{SFR}}$) of individual clouds is proportional to the dense gas mass surface density \citep{lada_2003, wu_2005, wu_2010, heiderman_2010, lada_2010, lada_2012, andre_2014, evans_2014}. However, to study individual molecular clouds in other galaxies, extragalactic surveys have historically focused on the brightest observable molecular emission lines: the low-$J$ transitions of $\rm{^{12}CO}$, which for simplicity we will refer to as CO. These transitions are sensitive to the total molecular gas mass, but cannot discriminate the gas mass in the densest regime. In order to probe the latter, less abundant molecules with transitions at higher critical densities ($n_\mathrm{crit}$) are needed. We refer to the definition of critical density from \cite{shirley_2015}. In this work, we will also make use of the most effective critical density defined in \citep{leroy_2017} (shortly the effective critical density${-}$hereafter $n_\mathrm{eff, crit}$). The effective critical density depends on a transition, kinetic temperature and optical depth. Molecules such as HCN, HNC and \HCOp{} have higher dipole moments than CO and its isotopologues and hence higher $n_\mathrm{eff, crit}$. Therefore, emission from these molecules has been used to probe the amount of denser molecular gas, which is more closely related to star formation than the lower density molecular medium traced by low-$J$ CO emission. The ratios between these lines and CO\Jone, in particular HCN/CO\Jone{}, are assumed to be a good proxy for the dense gas fraction.

In the seminal study by \cite{gao_2004a, gao_2004b}, they obtained galaxy integrated measurements of HCN and total infrared luminosities (TIR) to determine the dense gas mass and star formation rate, respectively. They found a linear relation between the HCN and TIR luminosities, therefore HCN\Jone{} emission appears to be directly correlated to the level of star formation activity. \cite{gao_2004a, gao_2004b} and other studies \citep{gao_2007, gracia_carpio_2008, krips_2008, juneau_2009, garcia_burillo_2012, privon_2015} investigated whole galaxies and their centres. These studies were focusing on the relation on global scales, i.e. averaging over different regions with different physical characteristics. However, studies of molecular lines other than CO within extragalactic sources are difficult. The emission of these molecules (HCN, \HCOp, HNC, etc) is typically very weak (e.g.\ HCN is ${\sim}20{-}30$ times weaker than CO \citealp{gao_2004b}), and for their detection more observing time is required.

Some other studies have investigated dense gas (HCN and \HCOp) in giant molecular associations in nearby galaxies: in M31 \cite{brouillet_2005}, M33 \cite{buchbender_2013}, and in the outer spiral arm of M51 \cite{chen_2017, querejeta_2019}. However, in order to understand the physics in galaxy discs, we need to resolve these in regions of faint molecular emission as well.

Over the last decade, many studies have focused on observing resolved galaxy discs in faint molecular lines. \cite{kepley_2014} mapped these lines in the starburst galaxy M82 using the Green Bank Telescope (GBT). These authors found that the HCN and \HCOp{} emission correlates with star formation and more diffuse molecular gas. \cite{usero_2015} targeted HCN emission across 60 regions within 30 nearby galaxies at kiloparsec resolution using the IRAM \mbox{30-m} telescope. This study investigated and found for the first time systematic variations in the dense gas fraction ($f_\mathrm{dense} = M_\mathrm{dense}/M_\mathrm{mol}$) traced by HCN/CO\Jone{}: higher values are seen in the centres of galaxies than in their outer parts. They also conclude that the star formation efficiency of dense molecular gas ($SFE_\mathrm{dense} = SFR/M_\mathrm{dense}$) traced by the TIR/HCN luminosity ratio is lower in the centres of galaxies than in the outer disc.

The recent `Emir Multiline Probe of the ISM Regulating galaxy Evolution' (EMPIRE) IRAM \mbox{30-m} EMIR survey, was the first survey to obtain a sensitive wide-area mapping of so-called denser molecular gas tracers (e.g. HCN) across the discs of nine star-forming galaxies at $30\arcsec$ resolution ($\sim\,1-2$\,kpc) \citep{bigiel_2016, jimenez19}.
Similarly, \cite{bigiel_2016} found that $SFE\mathrm{_{dense}}$ strongly depends on local environment in M51. Moreover, \cite{jimenez19} showed that these variations are present in the full sample of nine galaxy discs. \cite{gallagher_2018a} mapped high critical density molecules (HCN, \HCOp, HNC, CS) and CO isotopologues (\thCO, \CeiO) across four nearby galaxies at $8\arcsec$ or a few hundred parsec resolution using the Atacama Large Millimeter/$\allowbreak$submillimeter Array (ALMA). This study looked into the connection between the dense gas fraction, SFR and the local environment. An important result from the above mentioned studies was that while the fraction of denser gas increases towards the centres of galaxies, its efficiency to form stars is typically greatly reduced \citep{usero_2015, bigiel_2016, gallagher_2018a, jimenez19, jiang_2020}. This result agrees well with studies in the Milky Way where it appears that the dense gas fraction and the star formation efficiency of dense gas within the Central Molecular Zone are higher and lower, respectively, compared to local Milky Way clouds (e.g. \citealp{longmore_2013a, kruijssen_2014, barnes_2017}). The trend that has been seen between the TIR/HCN and HCN/CO in galactic centres, as shown in theoretical work by \citet{kruijssen_2014a}, and found in observations \citep[e.g.][]{jones_2012,longmore_2013a,usero_2015} supports the idea that there is no absolute density threshold for star formation and that the overdensity relative to the background is important (e.g. \citealp{federrath12}). For example, going towards the centre of the galaxy, more dense gas is present, which increases the HCN/CO ratio, and, hence, it is expected to also form stars at a higher efficiency. However, in the centre, the HCN-emitting gas is not tracing the relative overdensities, but rather the bulk dense gas, which is mostly not star-forming. The use of the HCN/CO ratio as a tracer for dense, star-forming gas in these regimes is then problematic (e.g. \citealp{bigiel_2016, jimenez19}).

Together, the studies by \citet{usero_2015, bigiel_2016, gallagher_2018a, jimenez19, jiang_2020} provide the first resolved view of dense molecular gas in galaxy discs. However, the resolution that they achieve in the extragalactic studies is still only ${\sim}500$\,pc to 2\,kpc. This is enough to resolve galaxy discs and distinguish central and disc regions, but it is still much bigger than the size of an individual molecular cloud (e.g. ${\sim}50{-}100$\,pc). As a result, observations like EMPIRE mix together many clouds in distinct evolutionary states (the typical distance between independent regions with distinct evolutionary states is ${\sim}100{-}200$~pc; see e.g.~\citealt{kruijssen_2019, chevance_2020a, kim_2020}) and physical environments \citep{hughes_2013, colombo_2014}. Recent decades and years have seen rapid growth in observations of CO \citep{heyer_2015}, which have been undertaken with single dish telescopes \citep[e.g.][]{yamaguchi_1999, dame_2001, regan_2001, moriguchi_2001, helfer_2003, garcia-burillo_2003, mizuno_fukui_2004, gratier_2012, burton_2013, barnes_2015}, as well as with the current generation of interferometer telescopes (e.g. NOEMA and ALMA) \citep[e.g.][]{engargiola_2003, rosolowsky_2003, rosolowsky_2007, sheth_2008, hirota_2011, schinnerer_2013, schruba_2017, sun_2018, egusa_2018, faesi_2018, leroy_2021b, maeda_2020, Rosolowsky_2021}. The big step forward in terms of sensitivity, resolution and sample size is the current PHANGS-ALMA survey (Physics at High Angular-resolution in Nearby GalaxieS with ALMA). The PHANGS-ALMA survey maps CO\Jtwo{} emission across a sample of 74 nearby star-forming galaxies with resolution high enough to detect individual GMCs across galaxies' discs \citep[PI: E.~Schinnerer;][]{leroy_2021b, Rosolowsky_2021}.

The next logical step is to also map the denser gas content of individual molecular clouds. The single dish studies have provided an insight into how the dense molecular gas is distributed in nearby galaxies \citep{kepley_2018, viaene_2018, watanabe_2019}. Moreover, despite the difficulties attaining sensitivity at high angular resolution, several works have begun to map high critical density lines down to cloud scales in nearby galaxies using interferometers (e.g. \citealp{viti_2014, murphy_2015, chen_2017, walter_2017, gallagher_2018b, querejeta_2019}). Observations of high critical density molecules at a high spatial resolution allows us to measure dense molecular gas as a function of cloud surface density, dynamical state, and evolutionary state. Several studies have already taken pioneering steps in this direction, mapping HCN and \HCOp{} emission at high physical resolution, and demonstrated the full potential of such observations despite the limited field of view, i.e., in chemical modelling \cite{viti_2014}, studying the outflows \cite{walter_2017}, investigating how the $SFE_\mathrm{dense}$ varies at 100\,pc scales \cite{querejeta_2019}. \cite{gallagher_2018b} combined high resolution CO measurements with lower resolution EMPIRE and ALMA (ACA) HCN maps to connect cloud properties to the dense gas fraction. However, a high resolution, sensitive, resolved HCN and \HCOp{} map across a large portion of a galaxy disc is still lacking within the literature.

In this work, we present new observations using the NOEMA interferometer targeting one galaxy: NGC~3627 (source information is listed in Table~\ref{tab:NGC3627_prop}). This survey currently represents the highest resolution observations ($1.85\arcsec = 102$~pc) across a large part of a galaxy disc using high critical density molecules (line properties are given in Table~\ref{tab:line_prop}). We use these observations to study the physical conditions of the denser gas at the scale of individual molecular clouds, to examine how the dense molecular gas is distributed across the galaxy's disc at these scales, and to study various density-sensitive line ratios. We investigate how dense molecular gas is linked to star formation at cloud scales. We also study various line ratios: the observed molecular lines relative to CO\Jtwo{} and the line ratios among the high critical density molecules (HCN, HNC and \HCOp). Investigating line intensities relative to CO\Jtwo{} emission, in particular HCN/CO\Jtwo{}, we are able to determine where the more dense molecular gas is present relative to the molecular gas content and how it is affected by the environment and star formation. Moreover, by studying the line ratios among the high critical density molecular lines such as HCN, HNC and \HCOp{}, we access the physical and chemical processes that set the cloud properties.

NGC~3627 is a nearby, star-forming galaxy with a strong bar, part of the M66 group (Leo Triplet) \citep{garcia_1993}. It is also classified as LINER/$\allowbreak$Seyfert type~2 galaxy \citep{ho_1997, filho_2000}. \cite{watanabe_2019} observed three regions in NGC~3627 (the centre, a~bar end, and a spiral arm) in 3~mm band using IRAM \mbox{30-m} and Nobeyama \mbox{45-m} telescopes. They detected ${\sim}10$ molecular species in each region, finding that the chemical composition is similar among these regions. NGC~3627 has been mapped in HCN as part of the EMPIRE survey using the IRAM \mbox{30-m} telescope \citep{jimenez19} and by ALMA \citep{gallagher_2018a}. \cite{murphy_2015} found a spatial offset between the peak intensity of HCN and \HCOp{} and tracers of recent star formation in the centre and the bar ends in NGC~3627. This study also found that the dynamical state of the gas plays a more important role in star formation than the abundance of the dense gas. \cite{beuther_2017} investigated dynamics in NGC~3627 bar ends, finding multiple velocity components in CO\Jtwo{} that originate from orbits coming from the bar and the spiral arm.

The paper is organized as follows. In Section~\ref{sec:observations}, we outline the reduction and imaging of the NOEMA observations. We summarise the ancillary observations used throughout this work in Section~\ref{sec:anc_obs}. In Section~\ref{sec:results}, we present the moment maps for each line, and discuss trends of the integrated intensity as a function of radius, star formation rate surface density, and molecular gas surface density, as well as dense gas velocity dispersion and various line ratios. In Section~\ref{sec:discussion}, the results of this work are discussed in the framework of our current understanding of dense molecular gas properties and star formation. Finally, in Section~\ref{sec:summary}, we summarise the findings of this work.

\begin{figure*}
	\includegraphics[scale = 1.5]{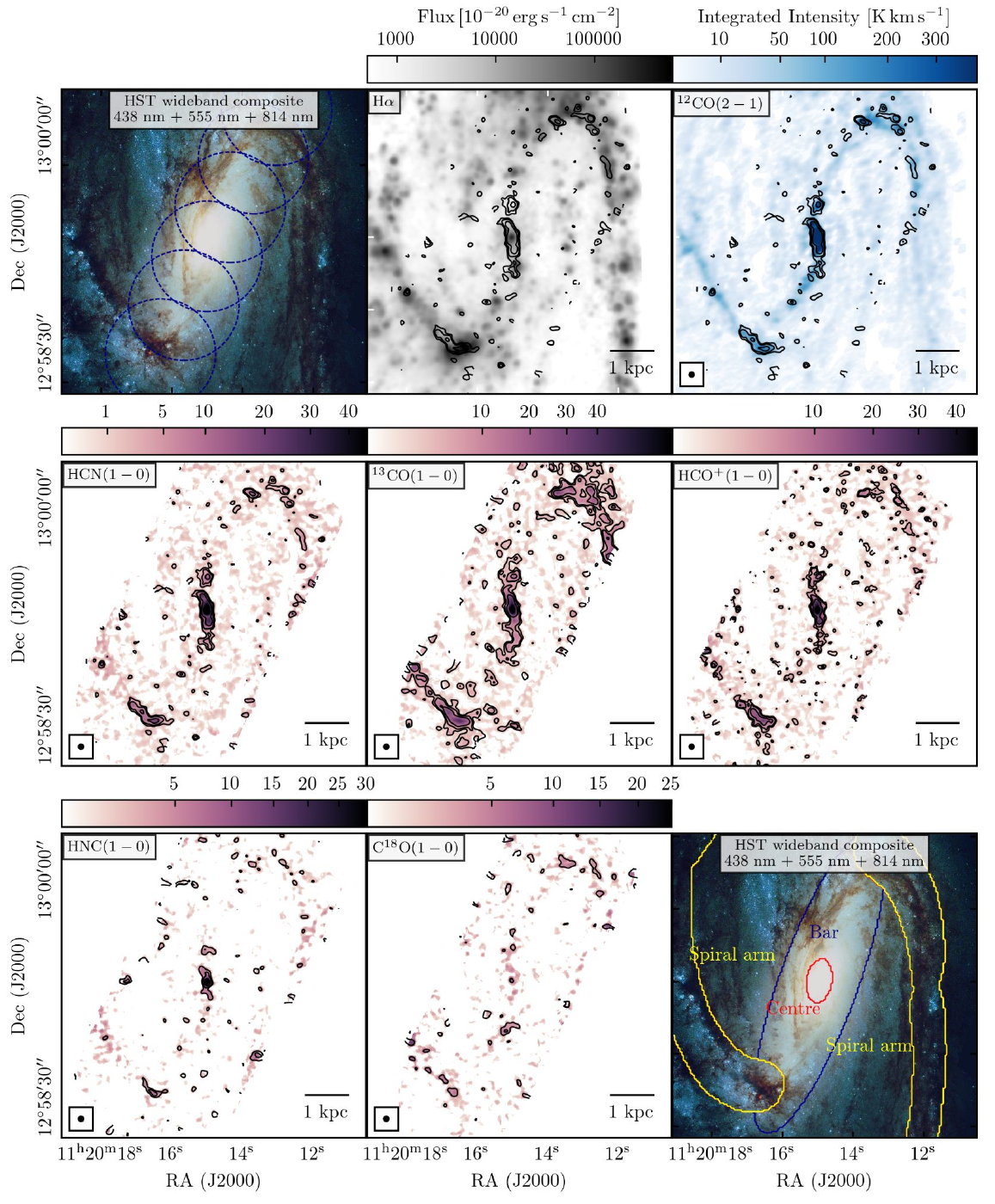}
    \caption{Top row: Pointing pattern of the NOEMA mosaic corresponding to a frequency of 109.93 GHz overlayed on the wide field composite image of NGC~3627 obtained with the Hubble Space Telescope (\citealp{lee_2021}, left panel), extinction-corrected H$\alpha$ map at $1.86\arcsec$ ($\approx$100~pc) resolution taken by the PHANGS-MUSE survey ${-}$ we show all the H$\alpha$ emission here, i.e. BPT and HII masking is applied to this image in a later step to recover actual SFR (see Section \ref{sec:anc_obs}, E.~Emsellem et al.\ in prep) (middle panel), CO\Jtwo{} integrated intensity map at $1.95\arcsec$ ($\approx$107~pc) taken by the PHANGS-ALMA survey (\citealp{leroy_2021b}, right panel). Contours show HCN(1-0) integrated intensity at 3, 5, and 10 $\sigma$ levels (see below). Middle row: HCN, \thCO{} and \HCOp{} integrated intensity map in $\mathrm{K\,km\,s^{-1}}$ observed by NOEMA at the working resolution of $1.95\arcsec$ ($\approx$107~pc) adopted in this work. Bottom row: Same as in the second row but for HNC and \CeiO{}. Contours in the middle and bottom row show levels of 3, 5, and 10 of signal/$\allowbreak$noise ratio for each line. The circle on the bottom left corner of the integrated intensity maps is the synthesised beam ($\mathrm{1.95\arcsec}$). Bottom right panel: Same as in the top-left panel with overlayed contours of environmental masks \citep{querejeta_2020}.}
    \label{fig1:new_data}
\end{figure*}


\GalaxyProp{}

\Ncrit{}

\section{IRAM observations, ancillary data and the production of integrated intensity maps}{\label{sec:observations}}

\subsection{IRAM observations}

The observations were taken at 3~mm using both the IRAM NOEMA interferometer at Plateau de~Bure and the IRAM \mbox{30-m} single dish. The 3~mm continuum emission was not detected. Since a multiplicative interferometer filters out the low spatial frequencies, i.e. spatially extended emission, we used the IRAM \mbox{30-m} observations to recover the low spatial frequency (`short- and zero-spacing') information missed by NOEMA at a depth that matches the interferometric data. In the following sections, we describe the observing strategy, calibration, joint imaging, and deconvolution processes. Table~\ref{tab:obs} summarises the parameters of the interferometric observations.

\subsubsection{Interferometric observations and calibration}

Interferometric measurements of NGC~3627 were obtained with NOEMA. The sideband separating receivers were tuned to observe from 86.9 to 94.6\GHz{} (lower sideband) and from 102.4 to 110.1\GHz{} (upper sideband). The \texttt{POLYFIX} correlator yielded a total bandwidth of $2 \times 7.7\GHz$ per polarization at a spectral channel spacing of 2\MHz. Each intermediate frequency baseband was further split into up to $16 \times 64\MHz$ chunks of high spectral channel spacing. These chunks were centred around potential lines inside the lower and upper sidebands. This yielded spectra with a 62.5\kHz{} channel spacing that we further smoothed to reach three different spectral resolutions: 5, 10, and 20\kms{}. In short, the frequency setup was chosen to simultaneously cover the $J=1{-}0$ lines of \thCO{} and \CeiO{} in the upper sideband, and the $J=1{-}0$ lines of \HCOp{}, HCN, and HNC in the lower sideband. Their frequencies are listed in Table~\ref{tab:line_prop}.

We observed a mosaic of 6~pointings aligned along the bar of NGC~3627 at a position angle of $-26\deg$. The neighbouring pointings were separated by $22.5\arcsec$, which is half the primary beam size at 110~GHz. The mosaic thus covers a roughly rectangular field of view of about $2.7\arcmin\times0.9\arcmin$ ($\mathrm{\approx1.5\,kpc\times0.5\,kpc}$). These measurements were carried out with 8 or 9 antennas in the C and A configurations (baselines from 15 to 750 meters) from February to May 2018. This amounts to 30.5 hours of telescope time (12~hours in~C and 18.5~hours in A~configurations). The on-source time is equal to 8.8~hours with a 9~antenna array. During the observations, the typical precipitable water vapour ranged from 1 to 3~mm in A~configuration and 3 to 8~mm in C~configuration. The typical system temperature was between 70 and 150\K, depending on the weather.

We used the standard algorithms implemented in the \texttt{GILDAS/$\allowbreak$CLIC} software to calibrate the NOEMA data.\footnote{See \texttt{http://www.iram.fr/IRAMFR/GILDAS} for more information on the GILDAS software~\citep{pety05}.} The radio-frequency bandpass was calibrated by observing the bright quasars 3C84 (${\sim}14$\Jy), 3C279 (${\sim}15$\Jy), and 0851$+$202 (${\sim}5$\Jy). Phase and amplitude temporal variations were calibrated by fitting spline polynomials through regular measurements of two nearby quasars (1222$+$216 with a flux of about 2.0\Jy{} at 17.5$\degr$ distance,
and 1116$+$128 with a flux of about 0.5\Jy{} at 0.5$\degr$ distance). One of the NOEMA secondary flux calibrators (either MWC349 or LKH$_{\alpha}\,101$) was observed during each track, which allowed us to improve the accuracy of the absolute flux scale of the interferometric data to ${\sim}10\%$.

\subsubsection{Joint imaging and deconvolution of the interferometric and single-dish data}

The single-dish observations were taken with the IRAM \mbox{30-m} telescope as part of the EMPIRE large program (PI: F.~Bigiel) from December 2014 to December 2016. \citet{jimenez19} describe in detail the observations and data reduction.

Following \citet{rodriguez08}, the \texttt{GILDAS/$\allowbreak$MAPPING} software and the single-dish data from the IRAM \mbox{30-m} were used to create the short-spacing visibilities not sampled by NOEMA using the \texttt{UV\_SHORT} task. In short, the \mbox{30-m} data cubes were first resampled around the redshifted frequencies observed with NOEMA and reprojected to the NOEMA phase centre. The data cubes were deconvolved from the IRAM \mbox{30-m} beam in the Fourier plane, and corrected for the NOEMA primary beam response in the image plane. After a last Fourier transform, pseudo-visibilities were sampled between 0 and 15~m, the difference between the diameters of the IRAM \mbox{30-m} and the \mbox{15-m} NOEMA antennas. These visibilities were then merged with the interferometric observations.

Each mosaic field was imaged and a dirty mosaic was built combining those fields in the following optimal way in terms of signal/$\allowbreak$noise ratio \citep{pety10}. The dirty cube is corrected for primary beam attenuation, which induces a spatially inhomogeneous noise level. In particular, noise strongly increases near the edges of the field of view. To limit this effect, both the primary beams and the resulting dirty mosaics are truncated. The standard level of truncation is set at 20\% of the maximum in \texttt{GILDAS/$\allowbreak$MAPPING}. The dirty image is deconvolved using the standard H\"ogbom CLEAN algorithm. CLEAN components were only searched for inside a mask produced from the EMPIRE \twCO{} \Jone{} cube obtained at the IRAM \mbox{30-m} telescope. Pixels within each channel with a signal/$\allowbreak$noise ratio of the CO\Jone{} line larger than~3 are included in this three-dimensional mask. This gives a shallow mask that loosely follows the galaxy velocity pattern as the angular resolution is $27\arcsec$ and the typical sensitivity is good $(\sim10\mK)$. The resulting data cubes are then scaled from Jy~beam$^{-1}$ to main beam brightness temperature ($T_\mathrm{b}$) scale using the synthesized beam size (see Table~\ref{tab:obs}). The channel width of the final data cubes is 5\kms. The resulting beam size, and corresponding noise values of the final combined maps used throughout this work are presented in Table~\ref{tab:obs}.

\TabObsNOEMA{}

\subsection{The production of integrated intensity maps}{\label{subsec:ii_maps}}

We produce integrated intensity maps from masked spectral cubes for each line. The mask is based on the ALMA CO\Jtwo{} cube (resolution $1.5\arcsec$), because we expect to detect HCN emission only in regions of the galaxy with detected CO emission \citep{jimenez19, querejeta_2019}. We convolve all NOEMA data cubes to a common working resolution of $1.95\arcsec\times1.95\arcsec\times5\kms$ ( $1.95\arcsec\approx$102\,pc). The ALMA CO\Jtwo{} cube (see the following section for more details) has been spectrally and spatially smoothed and regridded to match the NOEMA observations. 

We make use of an expanding masking technique in order to create masks for our data set \citep{pety_2013}. To construct the mask, we calculate the three-dimensional noise cube from signal-free parts of the spectrum. Then we define an initial mask where we select all pixels with signal/$\allowbreak$noise ratio higher than~4 over at least two neighbouring channels. The mask is then expanded to cover all the pixels defined by a lower~2 signal/$\allowbreak$noise ratio mask \citep{ros_ler_2006}.

We apply the CO\Jtwo{} mask to all of our line data cubes, and we determine the integrated intensity maps by summing intensities along the velocity axis for all lines of sight and multiplying by the channel width ($\mathrm{\Delta \upsilon = 5}$\kms). We construct a two dimensional noise map ($\Delta_\mathrm{rms}$) based on the signal-free parts of each spectrum. Then we estimate the uncertainty in the integrated intensity by scaling this value by the channel width $\mathrm{\Delta \upsilon}$ and the square root of the number of signal channels~$N$: 
\begin{equation}{\label{eq:uncertainty}}
    \Delta_\mathrm{I} = \Delta_\mathrm{rms} \Delta \upsilon \sqrt{N}.
\end{equation}
The signal/$\allowbreak$noise ratio is then calculated by dividing the integrated intensity map by this uncertainty map.

\subsection{Ancillary data}\label{sec:anc_obs}

Throughout this work, we make use of a set of ancillary observations to determine the molecular gas and star formation rate surface densities. These are summarised in this section. 

\subsubsection{PHANGS-ALMA CO\Jtwo}

As a tracer of the molecular gas surface density in NGC~3627, we make use of $\mathrm{CO\Jtwo}$ molecular line observations obtained with the ALMA interferometer. These were performed as part of the PHANGS-ALMA survey \citep[PI: E.~Schinnerer;][]{leroy_2021b}. Both \mbox{12-m} array, \mbox{7-m} array and the total power antennae (ACA) were used for the mosaic observations, therefore full spatial information is recovered for the whole CO disc of NGC~3627. The interferometric data were calibrated with the ALMA calibration pipelines and imaged with the PHANGS-ALMA pipeline \citep{leroy_2021a}. The total power data were calibrated following the method presented by \citet{herrera_2020}. Then, interferometric and total power data were aligned, combined with the CASA feather task, and post-processed to $T_\mathrm{b}$ cubes using the PHANGS-ALMA pipeline. Moment maps were also generated with this pipeline. The data we used in this work are from the internal data release version~$3.4$ processed with the PHANGS-ALMA pipeline version~1. We refer the reader to \cite{leroy_2021b} for more details about the data processing. Imaging was done using CASA version~$5.4.0$. After calibration and imaging, the data cube was convolved to produce a round beam. The typical rms noise in brightness temperature units is ${\sim}0.17$~K per 2.5\kms{} channel (see also \citealp{schinnerer_2019}). In our work, we use a CO\Jtwo{} data cube convolved to $1.95\arcsec$ resolution with the channel width of $5\kms$.

\subsubsection{PHANGS-MUSE H$\alpha$ data}{\label{sec:sfr}}

We determine the star formation rate in NGC~3627 from the $1.5\arcsec$ H$\alpha$ map observed with MUSE/VLT (the Multi-Unit Spectroscopic Explorer). These observations were performed as part of the PHANGS-MUSE survey. We make use of data from internal release version~$2.0$ (PI: E.~Schinnerer; E.~Emsellem et al.\ in prep.\ for full details of data reduction). To estimate the star formation rate at each pixel, we make use of the MUSE extinction-corrected H$\alpha$ map at $1.95\arcsec$ resolution. The extinction is calculated using the measured Balmer decrement assuming case~B recombination, i.e. an intrinsic H$\alpha$/H$\beta$ ratio of 2.86, which corresponds to a temperature of $10^4$~K and electron density of $100$~cm$^{-3}$ \citep{osterbrock_1993, dominguez_2013}. The recombination coefficients are robust to realistic changes in temperature and density \citep{osterbrock_1989}. We make use of a \citet{calzetti_2000} extinction curve, and for the extinction-corrected map, only H$\alpha$ and H$\beta$ with a signal/$\allowbreak$noise ratio better than~15 are used, whereas the values below this threshold are masked (see C.~Faesi et al.\ in prep.\ for full details on the extinction correction).

In addition to being produced in the nebulae ionised by massive young stars, H$\alpha$ photons can originate from a wide range of other ionizing sources: gas ionised by an AGN, older stellar populations, planetary nebulae (PNe), and supernovae. Thus simply adding up all H$\alpha$ emission could overestimate the star formation rate and therefore bias our results. In order to ensure that we only make use of the H$\alpha$ emission associated with star formation, we first match the \ion{H}{II} region catalogue of F.~Santoro et al.\ (in prep), which uses two algorithms for finding spatially resolved (HIIphot ${-}$ \citealt{Thilker2000}) and point-like HII regions (DAOStarFinder ${-}$ \citealt{stetson_1987}) to isolate H$\alpha$ emission associated with massive star formation. Then we mask our H$\alpha$ map. However, this is not enough, since H$\alpha$ photons within \ion{H}{II} regions can still in principle originate from sources other than star formation. We further apply a Baldwin, Phillips and Terlevich (BPT) \citep{baldwin_1981} cut with line luminosity (in units of solar luminosity${-}\,L_{\odot}$) ratio thresholds of $\log[\ion{N}{II}]/H\alpha < 0$ and $\log[\ion{O}{II}]/H\beta < -0.09$ \citep{baldwin_1981, kewley_2001, kauffmann_2003}. With these criteria, we remove all the pixels within the HII regions coming from ionizing sources other than the star-forming ones. All sight lines that do not satisfy these criteria are treated as a not a number (NaN). The \ion{H}{II} mask and the BPT cut remove ${\sim}40\%$ of the total H$\alpha$ flux from the initial H$\alpha$ map.

This map is then converted to the star formation rate surface density, $\Sigma_\mathrm{SFR}$, following \citet{calzetti_2007},
\begin{equation}
    \Sigma_\mathrm{SFR} = \frac{10^{-41.27} (3.08\times10^{21})^2 S_\mathrm{H\alpha} 4 \pi}{\Omega},
\end{equation}
where $\mathrm{\Omega}$ is the pixel angular area in steradians, $S_\mathrm{H\alpha}$ is the H$\alpha$ flux per pixel in units  of $\mathrm{erg\,s^{-1}\,cm^{-2}}$, and $(3.08\times10^{21})^2$ is the $\mathrm{kpc^2}$ to $\mathrm{cm^2}$ scaling factor. The factor of $\mathrm{5.37\times10^{-42}}$ has been calculated with assumptions that the H$\alpha$ has already been corrected for dust extinction, and a fully populated Kroupa initial mass function (IMF) \citep{kroupa_2001} taken over the range of stellar masses $0.1{-}100~\mathrm{M_{\odot}}$ \citep{murphy_2011, kennicutt_2012}. The use of H$\alpha$ emission provides an almost instantaneous measure of the star formation rate, tracing activity over the past ${\sim}0{-}10$~Myr (see \citealp{kennicutt_2012}). 

We note here several issues related to the conversions of H$\alpha$ emission to star formation rate, which are particularly relevant when assessing the star formation rate over small spatial scales (e.g. 100~pc scales). Firstly, the high sampling rate at high-resolution, can return regions with a low star formation rate ($\mathrm{\Sigma_{SFR}\,<\,10^{-3}\,M_{\odot}\,yr^{-1}\,kpc^{-2}}$), within which the IMF can become poorly sampled. The conversion from H$\alpha$ emission that assumes a fully sampled IMF may then under- or overestimate the star formation rate (e.g. \citealp{lee_2009, querejeta_2019}).
The IMF can be poorly sampled on smaller spatial scales where we probe either lower mass clusters or lower SFRs, which can cause some stochastic variation in the calculated SFR \citep{kennicutt_2012}.
Secondly, the H$\alpha$ line is subject to systematic uncertainties from dust attenuation and excitation variations in galaxies. Thirdly, the H$\alpha$ emission relies on the production of UV photons from massive ($>$10M$_{\odot}$) stars, which ionise the surrounding medium. Regions containing exclusively lower mass stars or which are well mixed with the diffuse ionised gas, are, therefore, not seen in H$\alpha$ emission, and not accounted for in our measurement of the SFR.


To estimate how much star formation is not traced by our extinction-corrected H$\alpha$ emission due to extinction, we compare to star formation rate estimates from a combination of far-ultraviolet (FUV) emission and mid-infrared (IR) emission. For that purpose, we use the $\mathrm{\Sigma_{SFR}}$ map from \cite{leroy_2019}. We use the $\mathrm{\Sigma_{SFR}}$ map determined from FUV+22\,\micron\ and convert to SFR by multiplying each value by the projected pixel surface area (0.075\,kpc$^2$ for a $5\arcsec\,{\times}\,5\arcsec$ pixel). We compare values coming within the radius of ${\sim}1$\,kpc from the centre of NGC~3627, and from the rest of the mapped region. For the central region, we measure a SFR of ${\sim}0.016\,\mathrm{M_{\odot}\,yr^{-1}}$ from the extinction-corrected H$\alpha$ emission and ${\sim}0.286\,\mathrm{M_{\odot}\,yr^{-1}}$ from FUV+22\,\micron. For the rest of the NGC~3627, the SFR traced by the extinction-corrected H$\alpha$ is ${\sim}1.27\,\mathrm{M_{\odot}\,yr^{-1}}$, whereas the SFR traced by FUV+22\,\micron\ is ${\sim}1.55\,\mathrm{M_{\odot}\,yr^{-1}}$. The large difference in SFR from the centre of NGC~3627 is due to the fact that H$\alpha$ emission associated with the AGN has been removed in the map used here, where it remains within the FUV+22\,\micron\ star formation rate estimate. Whereas, in the rest of the galaxy, where both the FUV+22\,\micron\ and H$\alpha$ emission are expected to better trace the star formation, we estimate that they are in agreement (i.e. within 17\,per cent). The small difference could be due to the high obscuration of the H$\alpha$ emission from deeply embedded star-forming regions, not corrected for by the Balmer decrement extinction correction, which then otherwise emit at longer wavelengths (i.e. as seen at 24\,\micron; e.g. see \citealp{kennicutt_2009, kim_2020}).

\section{Results}{\label{sec:results}}

\subsection{Integrated intensity maps}\label{sec:results_mom}

Figure~\ref{fig1:new_data} shows the ancillary data (first row) towards NGC~3627 (Section~\ref{sec:anc_obs}) and the integrated intensity maps (second row) for HCN, \thCO{} and \HCOp{}. We show the remainder of the detected lines, HNC and \CeiO{} in the bottom row of Figure~\ref{fig1:new_data}. The overlayed contours in the top row show the 3, 5, and 10\,$\sigma$ signal/$\allowbreak$noise ratio of HCN emission. The median uncertainty of the HCN integrated intensity across the mapped region is 2~K\kms. The corresponding $3\sigma$ sensitivity threshold for HCN is then 6~K\kms, which, assuming a CO\Jone/HCN ratio of 30 \citep{gao_2004b} and a CO\Jone-to-H$_2$ conversion factor of 1.2 $\mathrm{M_{\odot}\,(K\,km\,s^{-1}\,pc^{-2})^{-1}}$ calculated for NGC~3627 \citep{bolatto_2013, sandstrom_2013} implies a mass sensitivity of $\sim$216 $\mathrm{M_{\odot}\,pc^{-2}}$.

\begin{figure}
    \centering
	\includegraphics[width=0.8\columnwidth]{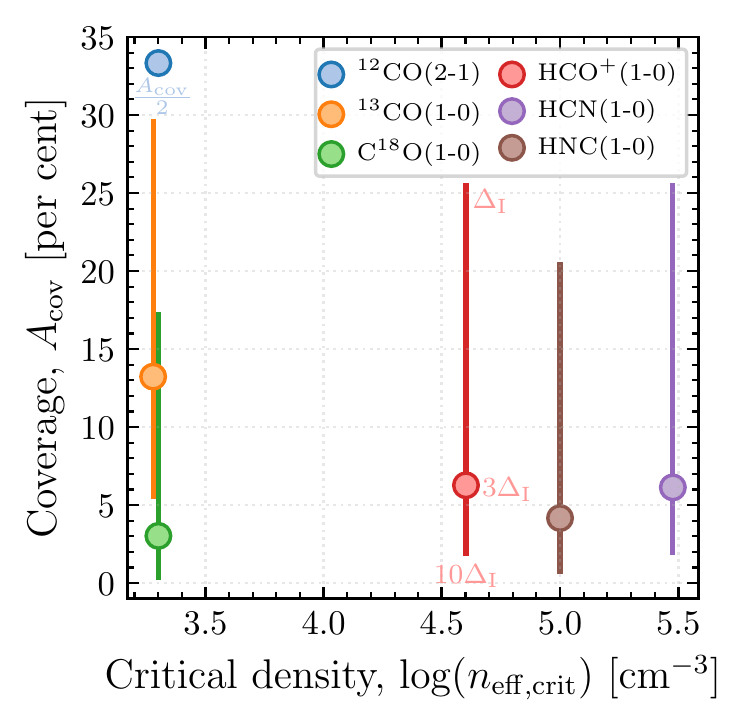}
    \caption{The percentage of pixels within the observed area that have significant values of the integrated intensity for each of the molecular lines, $A_\mathrm{cov}$. We define significant integrated intensity values as being higher than three times their associated uncertainty ($\Delta_\mathrm{I}$; see Section~\ref{subsec:ii_maps}). The error bars show the range of the percentage coverage for one and ten times the associated uncertainty. The variation between these limits for CO\Jtwo{} line is small, therefore the error bar is not visible on the plot. The molecular line transitions are ordered as a function of the effective critical density of emission. We note that we offset the position of the \thCO{} and \CeiO{} values on the x axis to avoid overlapping, due to their similar effective critical density (Table~\ref{tab:line_prop}). We also note that the coverage of the $^{12}$CO emission have be reduced by a factor of two for plotting.}
    \label{fig:ncrit_cov}
\end{figure}

In Figure~\ref{fig:ncrit_cov}, we show the fraction of the total observed area that has significant detections from each of the observed molecular lines as a function of their effective critical density listed in Table~\ref{tab:line_prop}. This area includes positions within the mosaic that have an integrated intensity value higher than three times the associated uncertainty (ranges for one and ten times the uncertainty are shown as error bars). In general, we find that the emission of high-critical density lines is much more spatially compact than the CO\Jtwo{} emission. The isotopologues of CO have a similar effective critical density, yet exhibit comparable coverage to the higher critical density lines. This can naturally be explained by the lower abundances of the CO isotopologues relative to $^{12}$CO, which causes weaker line emission that falls below our detection limit. Of the observed lines, we see that the \thCO{} line is the brightest and most spatially extended, whereas \CeiO{} is the faintest and most compact (see Figure~\ref{fig1:new_data}).

\subsection{Stacking procedure}{\label{sec:stacking_section}}

In order to recover emission from the low signal/$\allowbreak$noise lines of sight, we stack individual spectra as a function of galactocentric radius, star formation rate surface density and CO\Jtwo{} integrated intensity following \citet{schruba_2011}, \citet{caldu_primo_2016}, \citet{jimenez_2017b}, \citet{jimenez19}. To do this, first we convolve our data cubes and maps to a common angular and spectral resolution of $1.95\arcsec$ and 5\kms, respectively. Next, we regrid all data cubes and maps to a common world coordinate system (WCS). Last, we resample our line emission maps (including the ancillary data) onto the same hexagonal grid, where sampling points are spaced by half a beam size of the NOEMA observations (${\sim}1\arcsec$) thus oversampling the data by a factor of four.

Then we measure the average integrated intensity of each observed line in each bin. To do this, first we calculate the velocity at which the CO\Jtwo{} spectrum peaks within all the lines of sight. In the next step, we shift the spectra of each line of sight within the NOEMA and ALMA CO\Jtwo{} cubes by the velocity previously defined. This procedure results in cubes where all molecular line emission peaks at a velocity close to 0\kms. Averaging the shifted lines will increase the signal/$\allowbreak$noise ratio and by construction one knows where a (potentially weak) spectral line should build up. It should be highlighted, however, that this method relies on the robust detection of at least one bright line to determine the line of sight velocity by which all other spectra will be shifted. Typically, CO\Jtwo{} is the brightest line along each line of sight, so this is used as a prior. We exclude sight lines where CO\Jtwo{} is not robustly detected, since we are unlikely to detect the weaker line emission towards these positions.
 
We integrate the stacked spectrum to determine the average integrated intensity of the line. We use stacked CO\Jtwo{} spectra as a prior to determine the width of the integration window. First, we define the signal-free part of the spectrum. Next, we determine where the spectral line is defined using the masking technique described in Section~\ref{subsec:ii_maps}. We select all channels with signal/$\allowbreak$noise ratio above~4 and expand the mask to cover all neighbouring channels with signal/$\allowbreak$noise ratio above~2. Finally, this velocity window is used as a mask that we apply to the stacked spectra of the rest of the observed molecular lines. The integrated intensity is then calculated as a sum over the integration window multiplied by the channel width. The uncertainty of the integrated intensity is computed using the Equation \ref{eq:uncertainty}. Because we estimate the noise from the signal-free region of the stacked spectrum itself, this uncertainty properly accounts for any spatial oversampling, as well as for the potential spatial and spectral correlation of the data. This implies that the uncertainty of the integrated intensity varies across different sight lines and from tracer to tracer. We take the integrated intensity of the stacked spectrum if the following criteria are not satisfied. In cases when CO\Jtwo{} is detected, we measure the integrated intensity of the stacked spectrum, its peak, and the values in the two channels next to the peak. We then determine their S/N ratio. If either of the signal/$\allowbreak$noise ratios are below~3, the integrated intensity is taken as an upper limit of 3~times the uncertainty of the integrated intensity defined in Equation~\ref{eq:uncertainty}. In cases where CO\Jtwo{} is not detected (signal/$\allowbreak$noise ratio $<\,3$) and therefore the velocity window can not be determined, we calculate the integrated intensity as an upper limit of 3 times the uncertainty of the integrated intensity that is calculated as the rms noise of the line multiplied by a 30\kms{} velocity window. The stacked spectra and integration windows are shown in Figure~\ref{fig:stacked_spec_rad}, Figure~\ref{fig:stacked_spectra_co21} and Figure~\ref{fig:stacked_spectra_sfr} in the Appendix.

\subsection{Stacking results}

\begin{figure*}
	\includegraphics[scale=0.82]{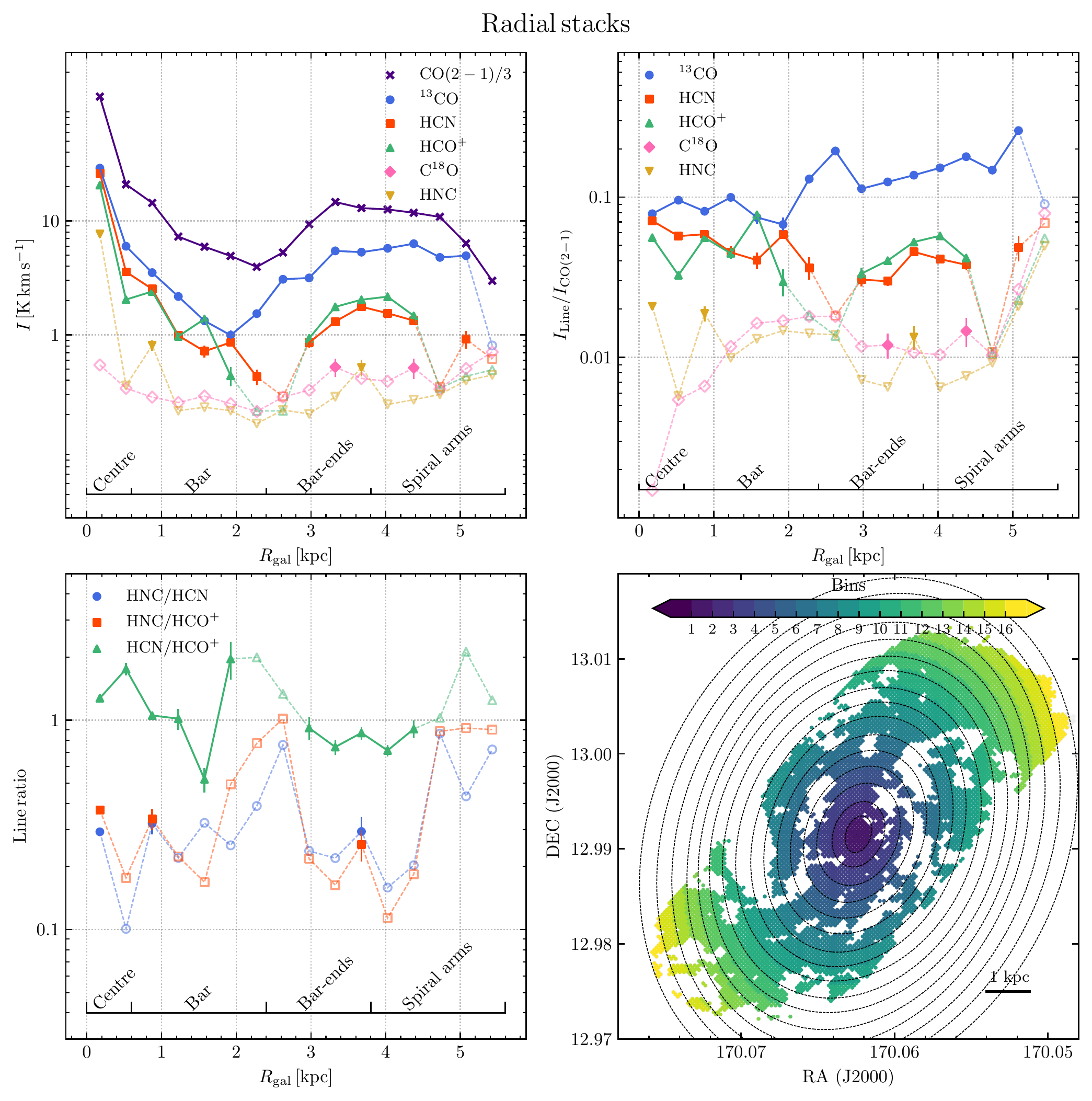}
    \caption{Stacked molecular line intensities binned by galactocentric radius. We show the integrated intensities of CO\Jtwo{}, CO isotopologues (\thCO{} and \CeiO) and the dense gas tracers in the top left panel and the line ratios of dense gas tracer per unit CO\Jtwo{} in the top right panel. Line ratios among the dense gas tracers are shown in the bottom left panel. All significant measurements (characterised as good and satisfactory detections) are connected with solid lines, whereas for non-detections we plot $3\Delta_\mathrm{I}$ values as an open symbol that are connected with dashed lines. Errorbars show one sigma uncertainties. Using the environmental mask, we label the location of each environment. The bottom right panel shows a map of galactocentric radius, where we colour-code each radial bin.}
    \label{fig:radial_profiles}
\end{figure*}

\begin{figure*}
	\includegraphics[scale=0.82]{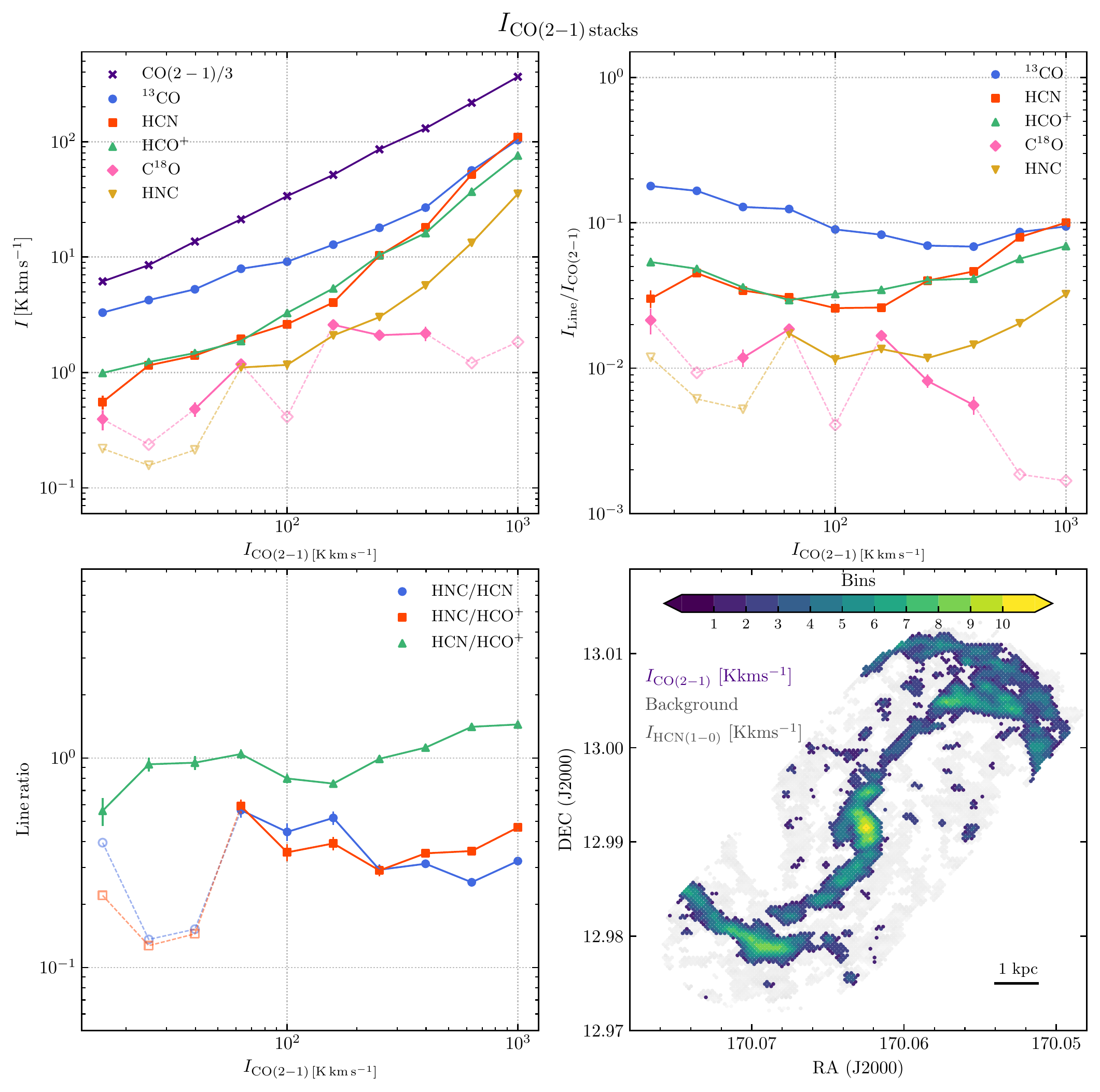}
    \caption{Same as Figure~\ref{fig:radial_profiles} but stacked by CO\Jtwo. In this case, we only use data points with signal/$\allowbreak$noise ratio $> 12$ in CO\Jtwo. We show the HCN integrated intensity map in grey in the bottom right panel. Coloured points represent the lines of sight that contribute to CO\Jtwo{} bins. }
    \label{fig:co_stacks}
\end{figure*}

\begin{figure*}
	\includegraphics[scale=0.82]{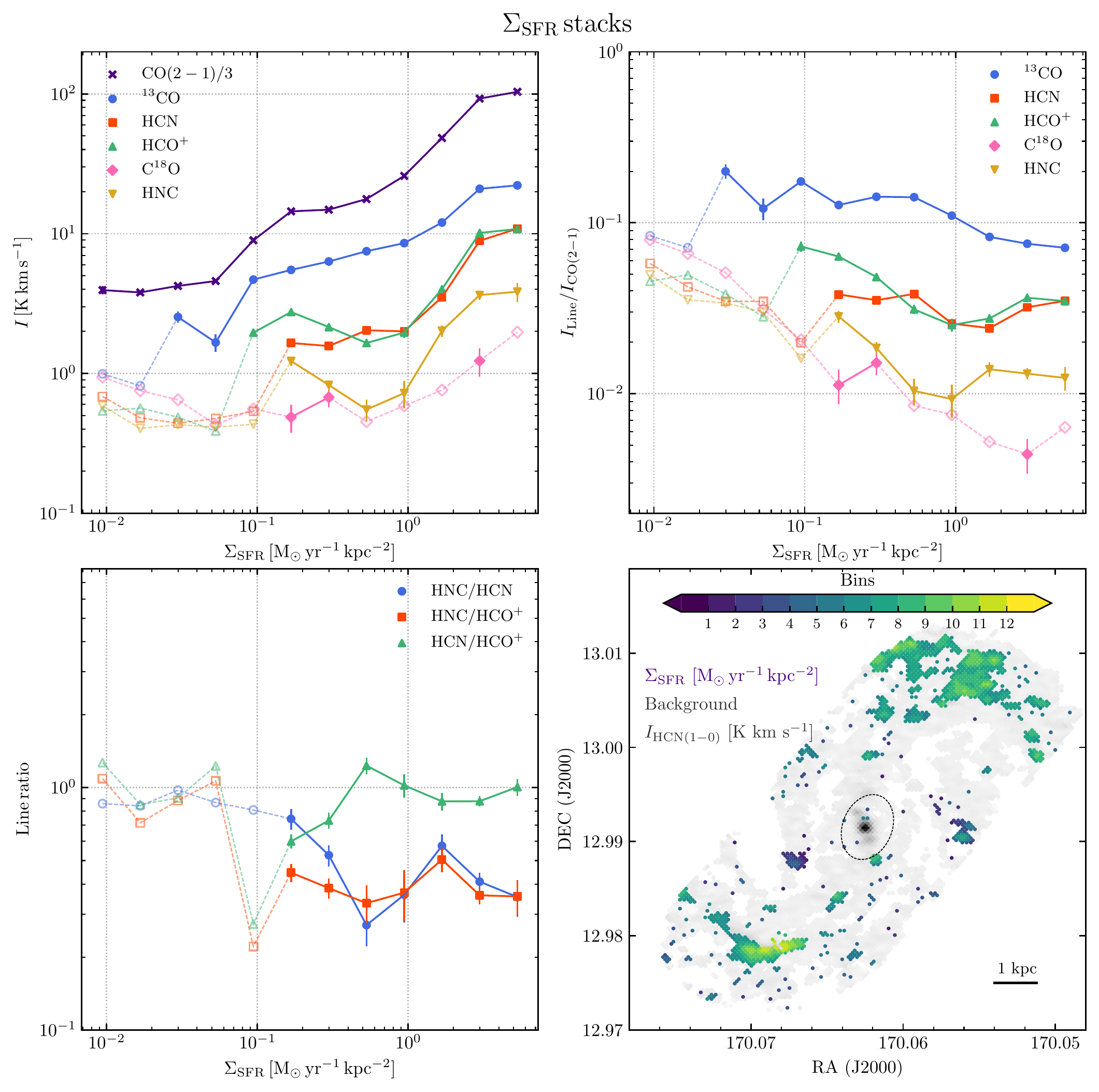}
    \caption{Same as Figure~\ref{fig:radial_profiles} but for stacks by $\mathrm{\Sigma_{SFR}}$. We stack data points from regions where star formation is traced by H$\alpha$ emission (Section~\ref{sec:anc_obs}). We exclude the centre (region within the circle with radius of $\sim\,600$ pc) where H$\alpha$ emission may not exclusively trace recent star formation. We also show the HCN integrated intensity map in grey in the bottom right panel.}
    \label{fig:sigma_sfr_stacks}
\end{figure*}

We stack the observed lines by the quantities measured at high spatial resolution: galactocentric radius, CO\Jtwo{} integrated intensity and SFR surface density. These stacked line profiles are shown in the top left panels in Figures \ref{fig:radial_profiles}, \ref{fig:co_stacks} and~\ref{fig:sigma_sfr_stacks}. 

We define radial bins in linear space, using a radial bin size of 350\,pc, which is approximately 3~times the beam size along the beam major axis. For stacking by CO\Jtwo{} integrated intensity, we use data points with a signal/$\allowbreak$noise ratio in CO\Jtwo{} integrated intensity greater than~12 due to lack of the emission of dense molecular tracers at fainter sight lines. This threshold selects $70\%$ of data points that contain bright CO\Jtwo{} emission. To stack by CO\Jtwo{} and $\mathrm{\Sigma_{SFR}}$, we define bins with widths of $\mathrm{10^{0.2}\,[K\,km\,s^{-1}]}$ and $\mathrm{10^{0.25}\,[M_{\odot}\,yr^{-1}\,kpc^{-2}]}$, respectively, in logarithmic space. To illustrate which parts of the galaxy entered which specific bin, we colour-code data points that contribute to the same bin and show them in the bottom right panels of Figures \ref{fig:radial_profiles}, \ref{fig:co_stacks} and~\ref{fig:sigma_sfr_stacks}. The CO\Jtwo{} bins are constructed in a way that the two brightest CO\Jtwo{} bins contain the very central part of NGC~3627, the following bins contain sight lines from the bar ends, and finally, the remaining bins contain fainter CO\Jtwo{} sight lines located along the spiral arms and the outskirts of the central region, bar, and the bar ends. The mid-high CO\Jtwo{} bins are associated with a few largest $\Sigma_\mathrm{SFR}$ bins.


The galactocentric radius stacks are shown in the top left panel of Figure~\ref{fig:radial_profiles}. In this figure, we label where different environments are located. Overall, all lines show the strongest emission in the centre of NGC~3627, after which their emission steadily decreases towards the bar where it reaches its minimum around 2\,kpc (except for \CeiO{} and HNC, where we do not recover emission). Line integrated intensities then increase towards the bar ends. The emission from the \thCO{} line is recovered in almost all bins, up to ${\sim}5$\,kpc. The \thCO{} intensity increases towards the centre and the bar ends, but the centre appears to be brighter than the bar ends. \cite{cormier_2018} reported the opposite (i.e. brighter \thCO{} in the bar ends than in the centre), but it was noted that this might be due to the low resolution at which the line was observed (${\sim}1.5$~kpc, compared to ${\sim}100$~pc in this work). Figure~\ref{fig:radial_profiles} shows that the bright \thCO{} at the bar ends covers a larger radial range than the emission at the centre. 

HCN and \HCOp{} emission are recovered along the bar (${\sim}2$~kpc). We find that HCN and \HCOp{} have similar integrated intensities in the centre (26 and 21~K\kms{}, respectively) and across the disc of NGC~3627. This was also shown for the inner ${\sim}4$~kpc region in NGC~3627 by \citet{gallagher_2018b} and \citet{jimenez19}. Furthermore, we see bright and constant HCN and \HCOp{} emission across the bar ends (${\sim}3{-}4$~kpc), where \HCOp{} is slightly brighter than HCN. Similar emission is seen in HNC, but HNC is overall fainter than HCN by a factor of $2{-}3$. \CeiO{} is the faintest in the centre in comparison with other lines from our sample. \cite{jimenez19} reported similar results for HNC and \CeiO{} in this galaxy. We do not see a clear trend for \CeiO{} and HNC, given their emission is only significant in the bar end and in the centre and bar end, respectively (see discussion in Section~\ref{sec:discussion}).


In the top left panel of Figure~\ref{fig:co_stacks}, we show the line integrated intensities as a function of CO\Jtwo{} emission. The integrated intensity of all lines increases with increasing CO\Jtwo{} integrated intensity. Taken at face value, this result implies that when more molecular gas is present, there is also more dense molecular gas. The brightest lines in our data set, \thCO\Jone, HCN and \HCOp, are recovered in all bins. We find that HCN and \HCOp{} emission is considerably weaker in the lower CO\Jtwo{} bins in comparison with \thCO{}. Across all the CO\Jtwo{} bins, HCN and \HCOp{} show similar integrated intensities. \HCOp{} emission is brighter than HCN for CO\Jtwo{} integrated intensities less than 200~K\kms{}, whereas for the higher CO\Jtwo{} values we find the opposite. At the very brightest CO\Jtwo{} bin, HCN shows the brightest emission (${\sim}110\,\mathrm{K}\kms$), followed by \thCO{} and \HCOp{} (103 and 76~K\kms, respectively). HNC emission is recovered in almost all the CO\Jtwo{} bins but is weaker than HCN and \HCOp{}. Finally, we recovered the emission of the faintest line in our data set, \CeiO{}, in half of the CO\Jtwo{} bins.


In the top left of Figure~\ref{fig:sigma_sfr_stacks} we show line integrated intensities as stacked by SFR surface density. The emission of the stacked lines is recovered for about half of the $\mathrm{\Sigma_{SFR}}$ bins. We note that the central region (where the observed molecular lines peak) is excluded for analysis in this case due to the H$\alpha$ emission not being sensitive to the presence of an embedded star formation present in this region (see Section \ref{sec:anc_obs}). We find that at $\mathrm{\Sigma_{SFR}}$ values of $\mathrm{10^{-1}}$ to $\mathrm{1\,M_{\odot}\,yr^{-1}\,kpc^{-2}}$ all the line intensities of denser molecular gas tracers are approximately flat, i.e. vary by a factor of $\le\,$2. This behaviour is different from the one seen when stacking by CO\Jtwo{} integrated intensity (Figure~\ref{fig:co_stacks}). Overall, line intensities show a higher correlation with CO\Jtwo{} than with $\mathrm{\Sigma_{SFR}}$. In the bottom right panel of Figure~\ref{fig:sigma_sfr_stacks}, where we show a map of the SFR surface density plotted over the HCN integrated intensity map, we see that HCN emission is also present in regions where there is no star formation traced by H$\alpha$ emission. $24\%$ of the total HCN (and $27\%$ of the CO\Jtwo) flux is present in regions with star formation traced by H$\alpha$ emission.

\thCO{} has overall the highest integrated intensity compared to the other observed 3~mm lines. For $\mathrm{\Sigma_{SFR}}$ of $5.3\,\mathrm{M_{\odot}\,yr^{-1}\,kpc^{-2}}$, \thCO{} has an integrated intensity of 22~K\kms{}. The HCN and \HCOp{} integrated intensities are a factor of ${\sim}2$ lower than the \thCO{} (both ${\sim}11$~K\kms). Figure~\ref{fig:sigma_sfr_stacks} shows that the HCN integrated intensity is approximately constant over $\mathrm{1\,dex}$ in $\mathrm{\Sigma_{SFR}}$, whereas for $\mathrm{\Sigma_{SFR}}$ greater than $\mathrm{1\,M_{\odot}\,yr^{-1}\,kpc^{-2}}$, HCN intensity increases by ${\sim}0.6$~dex. HNC emission appears to decrease from $0.2$ to $1\,\mathrm{M_{\odot}\,yr^{-1}\,kpc^{-2}}$, but its emission increases at $1\,\mathrm{M_{\odot}\,yr^{-1}\,kpc^{-2}}$ as the rest of the lines. The increasing trend of dense gas tracers with higher values of $\mathrm{\Sigma_{SFR}}$ we found is in agreement with \cite{gallagher_2018a} who showed that the denser gas mass traced by HCN correlates with SFR (in their work traced by H$\alpha$ and 24\,$\micron$ emission) in NGC~3627. Here we report similar results for the \CeiO{} line as we have done in previous paragraphs. \CeiO{} is the faintest line shown in Figure~\ref{fig:sigma_sfr_stacks}, with emission only recovered in three $\mathrm{\Sigma_{SFR}}$ bins.

\subsection{Comparison of the CO and HCN velocity dispersion}{\label{sec:line_widhts}}

At low spatial resolution ($\gg 100$\,pc), relative motions of the molecular gas caused by the galactic potential are thought to be the dominant factor broadening molecular line profiles. Hence, it is not possible to measure the intrinsic velocity dispersion of cold, denser clumps caused by their internal dynamics (e.g. turbulence and thermal motions). However, at scales of ${\sim}100$\,pc, we probe smaller clumps. Giant molecular clouds are also thought to decouple from the galactic dynamical environment at these spatial scales (e.g. \citealp{meidt_2020a, chevance_2020a}), and, hence, with high spatial resolution observations we might be able to measure these intrinsic properties.

\begin{figure}
	\includegraphics[width=\columnwidth]{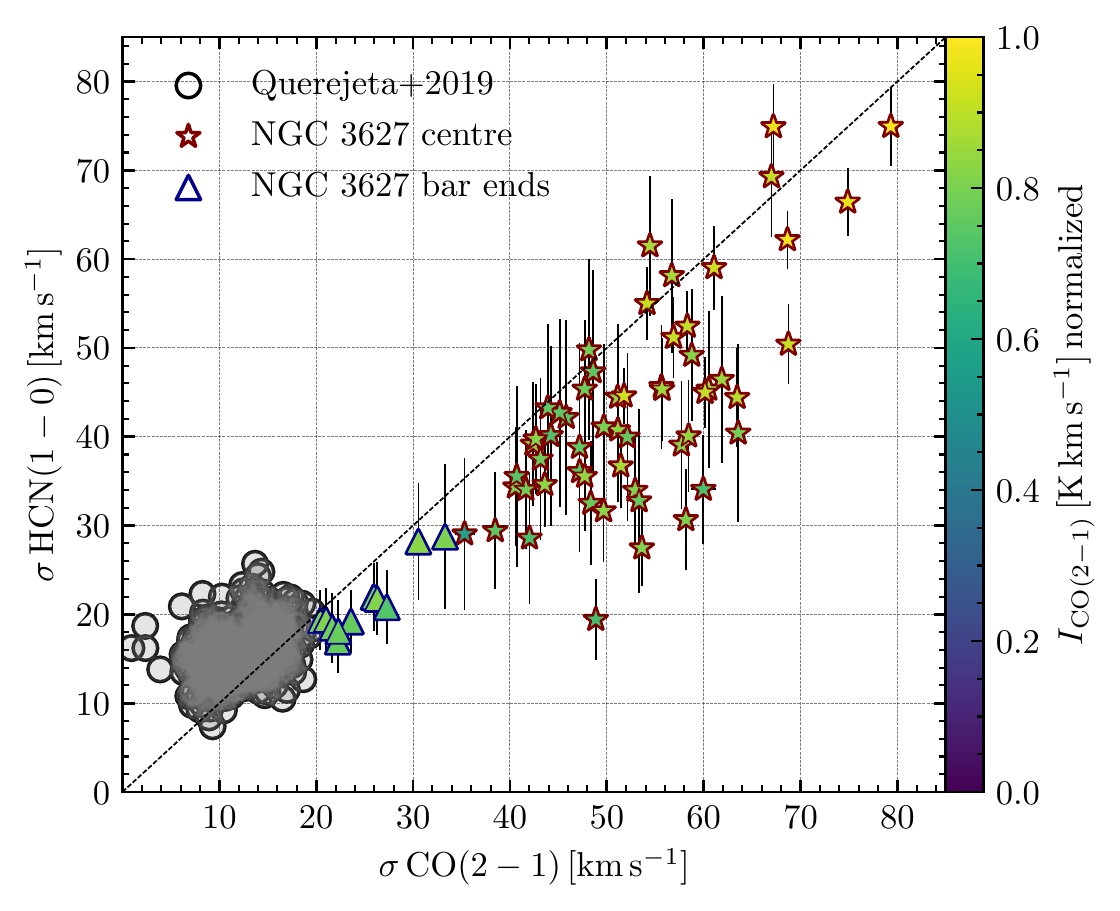}
    \caption{HCN velocity dispersion in comparison with CO\Jtwo{} velocity dispersion. We show velocity dispersion measurements from \citealp{querejeta_2019} as grey circles, whereas coloured points are sight lines from NGC~3627 colour-coded according to their CO\Jtwo{} integrated intensity. The dashed line shows where the points would lie in case of equal velocity dispersion in HCN and CO\Jtwo{} emission. The points from the centre of NGC~3627 are marked as stars with a red outline, whereas points coming from the bar ends are labelled as blue triangles. }
    \label{fig:linewidth_co21}
\end{figure}

Here we present the velocity dispersion of the HCN line emission, which we compare to the velocity dispersion of CO\Jtwo{} emission. We conduct the following method to measure the velocity dispersions of these molecular lines. The velocity dispersion calculation is based on the `effective width' approach \citep{heyer_2001, leroy_2016, leroy_2017a, sun_2018, querejeta_2019}, in which the velocity dispersion is computed as
\begin{equation}{\label{eq:line_width}}
    \sigma_\mathrm{Line} = \dfrac{I_\mathrm{Line}}{T_\mathrm{peak}\sqrt{2\pi}},
\end{equation}
where $I_\mathrm{Line}$ is the line integrated intensity in K\kms, and $T_\mathrm{peak}$ is the peak brightness temperature of the spectrum in units of~K. The uncertainty of the velocity dispersion is calculated by propagating the corresponding uncertainties. The velocity dispersion is determined within lines of sight with signal/$\allowbreak$noise ratio greater than~7 in both HCN and CO\Jtwo. We show our results in Figure~\ref{fig:linewidth_co21} where we colour-code our points according to their CO\Jtwo{} integrated intensity. We see that HCN line profiles become broader with brighter CO\Jtwo{} emission. This is also the case for CO\Jtwo, which was previously highlighted by \cite{sun_2018} and \cite{sun_2020}. Here we also compare our results with velocity dispersion measures within the literature.

Several works within the literature have also made a similar comparison to that presented in this section, albeit over various spatial scales using fundamentally different observations and methods to determine the velocity dispersion. Hence, a direct comparison is difficult and should be taken with caution. \cite{anderson_2014} investigated the velocity dispersions of dense molecular clumps at 1.45\,pc resolution in the 30~Doradus region within the Large Magellanic Cloud. Their result is that HCN and \HCOp{} velocity dispersion of molecular clumps in 30~Doradus are comparable with CO velocity dispersions, concluding that the denser molecular gas is not dynamically decoupled from the bulk molecular gas. They also found a trend of increasing clump brightness with increasing velocity dispersion. \cite{jimenez_2017b} measured HCN, \HCOp{} and \thCO{} line widths in six galaxies, including NGC~3627, at 1~kpc resolution from a Gaussian fit of radially stacked spectra. They found narrower line widths in the discs of their galaxies than in their centres. In the case of NGC~3627, the reason for the broad line widths in the central region is that there are unresolved rotational motions in the centre, as well as the motions along the spiral arms and the bar.

\citet{querejeta_2019} investigated the velocity dispersions of dense gas tracers in M51 over spatial scales (${\sim}100$\,pc) similar to those in our work. Importantly, this study used the same approach when calculating the `effective' velocity dispersion, therefore we can directly compare our results with \cite{querejeta_2019}. We show points from \cite{querejeta_2019} as grey circles in Figure~\ref{fig:linewidth_co21}. The velocity dispersions seen for sight lines coming from NGC~3627 are higher than those seen in M51 in both HCN and CO\Jtwo. There seems to be a consistent picture, where broader profiles of HCN correlate with broader and brighter CO profiles, which is then suggestive that we probe multiple clouds along the line of sight. Thus our results are also in agreement with \cite{querejeta_2019} and the studies mentioned in the paragraph above.

In Figure~\ref{fig:linewidth_co21} we see that there is a deviation from the ${1\,:\,1}$ relation as the HCN velocity dispersion is lower than the velocity dispersion of CO\Jtwo{} emission. This behaviour likely arises due to the HCN emission being less spatially extended than the CO\Jtwo{} (see Section~\ref{sec:results_mom} and Figure~\ref{fig:ncrit_cov}), which is expected also to be seen in velocity space, or due to the presence of more gas traced by CO\Jtwo. Larger velocity dispersion can result from the presence of complex line profiles. We see only a few sight lines where velocity dispersions are higher in HCN than in CO\Jtwo{} and these come from the sight lines with the brightest CO\Jtwo{} emission in NGC~3627.

To compare velocity dispersion across different regions in NGC~3627, we label points coming from the centre of NGC~3627 as stars with a red outline, whereas the sight lines coming from the bar ends are marked as triangles with a blue outline on Figure~\ref{fig:linewidth_co21}. Here we indeed see that velocity dispersions in the centre are higher than those from the bar ends. The deviation seen in the HCN-CO\Jtwo{} velocity dispersion in the centre continues further from the centre.

There could be several explanations for these systematic differences in the HCN line profiles relative to the CO emission. Towards the centre and bright star-forming regions infrared pumping of HCN could also broaden the velocity dispersion relative to CO \citep{matsushita_2015}. Another explanation is that HCN and CO\Jtwo{} may populate different orbits. CO\Jtwo{} line profile exhibit multiple velocity components across some sight lines which can broaden the line. For complex line profiles, it is possible that HCN is not tracing all the velocity components in CO\Jtwo. The main difference in velocity dispersion seen in the centre and the bar ends is that in the very centre HCN and CO\Jtwo{} trace the same molecular gas, i.e. the mean gas density in the centre might be higher than the HCN and CO\Jtwo{} effective critical densities (Table~\ref{tab:line_prop}). In the bar ends, however, the mean gas density is lower than the HCN effective critical density, therefore, HCN and CO\Jtwo{} do not trace the same gas. In conclusion, the HCN velocity dispersion relative to CO\Jtwo{} can be an identifier for complex line profiles. The importance of investigating each velocity component was demonstrated in \cite{henshaw_2020}. We expand this analysis later, where we link the presence of complex line profiles in CO\Jtwo{} with HCN emission in bar ends (Section~\ref{sec:dynamics}).

\subsection{Line ratios}{\label{sec:line_ratios}}

In this section, we present the results for different velocity-integrated brightness temperature line ratios (hereafter line ratios). Investigating the line ratios, we are able to access the physics and chemistry that describes molecular gas better than investigating line intensities themselves. We derived these line ratios from the spectral stacking procedure, for bins of CO\Jtwo{} integrated intensity, $\mathrm{\Sigma_{SFR}}$ and galactocentric radius (see Section~\ref{sec:stacking_section}). The error on the line ratio is calculated by the propagation of the uncertainties for the respective integrated line intensities,
\begin{equation}
    \mathrm{\Delta_{Line1/Line2}} = \left | \dfrac{I_\mathrm{Line1}}{I_\mathrm{Line2}} \right | \sqrt{\left( \dfrac{\Delta_\mathrm{Line1}}{I_\mathrm{Line1}}\right)^2 + \left(\dfrac{\Delta_\mathrm{Line2}}{I_\mathrm{Line2}}\right)^2},
\end{equation}
where $I_\mathrm{Line_i}$ is a given integrated line intensity and $\mathrm{\Delta_{Line_i}}$ is its uncertainty within each bin (their computation is described in Section~\ref{sec:stacking_section}), both in units of K\kms{}. The computed line ratio and the associated uncertainty are non-dimensional. 

In the following, we focus in more detail on the ratios of lines relatively to CO\Jtwo{} and the ratios between denser gas tracers (HNC/HCN, HCN/\HCOp{} and HNC/\HCOp{}).

\subsubsection{Line ratios with respect to CO\Jtwo{} integrated intensity}

We examine how the line intensities over the CO\Jtwo{} integrated intensity vary as a function of morphology across NGC~3627, CO\Jtwo{} emission, where the CO\Jtwo{} is assumed to trace the cloud-scale molecular surface density, $\mathrm{\Sigma_{mol}}$. We also look at how the ratio of the line intensities over the CO\Jtwo{} varies as a function of star formation rate surface density, $\mathrm{\Sigma_{SFR}}$. These line ratios are shown in the top right panels in Figures~\ref{fig:radial_profiles}, \ref{fig:co_stacks}, and~\ref{fig:sigma_sfr_stacks}.

First, we discuss the line ratios of the CO isotopologues. In Figure~\ref{fig:radial_profiles}, we show line intensities over the \mbox{CO\Jtwo} integrated intensity as a function of the galactocentric radius. We find that the \thCO/\mbox{CO\Jtwo} line ratio has the highest values across NGC~3627, with average values of $0.129\pm0.005$. We see that \thCO/\mbox{CO\Jtwo} is lower in the centre and inner bar, than compared to the outer bar region and the bar ends. The variation of \thCO/\mbox{CO\Jtwo} is $0.6$\,dex across a radial range of ${\sim}5$\,kpc.

In the top right panel of Figure~\ref{fig:co_stacks}, we show the line ratios as a function of \mbox{CO\Jtwo} integrated intensity. The \thCO/\mbox{CO\Jtwo} ratio appears to slightly decrease with increasing \mbox{CO\Jtwo} integrated intensity. The average \thCO/\mbox{CO\Jtwo} ratio over the \mbox{CO\Jtwo} bins is $0.109\pm0.002$. We find that \thCO/\mbox{CO\Jtwo} varies over $0.4$\,dex, whereas the \mbox{CO\Jtwo} integrated intensity varies by 2\,dex.

Similarly, we find in the top right panel of the Figure~\ref{fig:sigma_sfr_stacks} that the \thCO\Jone/\mbox{CO\Jtwo} ratio decreases with increasing $\mathrm{\Sigma_{SFR}}$. The average \thCO/\mbox{CO\Jtwo} line ratio is $0.126\pm0.006$ and this line ratio varies over $0.4$\,dex whilst $\mathrm{\Sigma_{SFR}}$ varies over ${\sim}3$ orders of magnitude.

The \CeiO/\mbox{CO\Jtwo} line ratio has the lowest values compared to the remaining line ratios we show here and also the lowest number of significant points. We find a value for \CeiO/\mbox{CO\Jtwo} of ${(12\,{\pm}\,2)}\times10^{-3}$ in the bar end of NGC~3627, and in the spiral arms it is ${(14\,{\pm}\,3)}\times10^{-3}$ (top right panel in Figure~\ref{fig:radial_profiles}). As a function of \mbox{CO\Jtwo} and $\mathrm{\Sigma_{SFR}}$, the mean \CeiO/CO\Jtwo{} ratio is $0.014\pm0.002$.

We now discuss the ratios of the remaining line integrated intensities over \mbox{CO\Jtwo} emission. These line ratios are used as a proxy of the dense gas fraction (e.g. HCN/CO). The HCN/\mbox{CO\Jtwo} integrated intensity can also be used to describe the dense gas fraction, $f_\mathrm{dense}$, since the \mbox{CO\Jtwo}/\mbox{CO\Jone} is well studied \citep{sandstrom_2013, law_2018, denbrok_2021}.

As a function of the environment (the top right panel of Figure~\ref{fig:radial_profiles}), HCN/\mbox{CO\Jtwo} line ratio has the highest measured value of $0.071\pm0.001$ in the centre of NGC~3627, after which slightly decreases towards the bar. The average line ratio across NGC~3627 is $0.046\pm0.003$. HCN/\mbox{CO\Jtwo} in the central region is ${\sim}2$ times higher than in the bar ends. From the top right panel in Figure~\ref{fig:co_stacks}, we report mean HCN/\mbox{CO\Jtwo} line ratio of $0.046\pm0.001$. The biggest value of $0.100\pm0.004$, in this case, is found in the very centre of NGC~3627. 

In case of CO\Jtwo{} bins (top right panel in Figure~\ref{fig:co_stacks}), it appears that HCN/\mbox{CO\Jtwo} has two regimes where the intersecting point is at the CO\Jtwo{} integrated intensity of $100\,{\rm K}\kms$. We find HCN/\mbox{CO\Jtwo} is almost constant in the first regime where the CO\Jtwo{} integrated intensity changes by an order of magnitude. In the second regime we find that the HCN/\mbox{CO\Jtwo} ratio varies by over $0.6\,\mathrm{dex}$. In the top right panel in Figure~\ref{fig:sigma_sfr_stacks}, HCN/\mbox{CO\Jtwo} ratio has significant points from $\Sigma_\mathrm{SFR}$ values of $10^{-1}\,\mathrm{M_{\odot}\,yr^{-1}\,kpc^{-2}}$. Here HCN/\mbox{CO\Jtwo} varies over $0.2\,\mathrm{dex}$ and the average line ratio is $0.035\pm0.002$. The maximum value of $0.038\pm0.002$ is found in the lowest recoverable $\Sigma_\mathrm{SFR}$ bin, while in the very last $\Sigma_\mathrm{SFR}$ we report a slightly lower value of this line ratio.

The rest of the line ratios (\HCOp/\mbox{CO\Jtwo} and HNC/\mbox{CO\Jtwo}) show a similar behaviour as the HCN/\mbox{CO\Jtwo}. \HCOp/\mbox{CO\Jtwo} and HNC/\mbox{CO\Jtwo} have average values of $0.047\pm0.003$ and $0.018\pm0.002$, respectively, across NGC~3627. We find higher values of HCN/\mbox{CO\Jtwo} in the centre of NGC~3627 than the \HCOp/\mbox{CO\Jtwo}, whereas it becomes vice versa in the bar ends. At the high-intensity end HCN/CO\Jtwo{} becomes overluminous compared to \HCOp{}/CO\Jtwo{} (top right panel in Figures~\ref{fig:radial_profiles} and~\ref{fig:co_stacks}). As a function of CO\Jtwo{} emission (top right panel in Figure~\ref{fig:co_stacks}), both the \HCOp/\mbox{CO\Jtwo} and the HNC/\mbox{CO\Jtwo} vary over $0.4\,\mathrm{dex}$. The average line ratios are $0.044\pm0.001$ for the \HCOp/\mbox{CO\Jtwo} and $0.017\pm0.001$ for the HNC/\mbox{CO\Jtwo}. We report similar variation of these two line ratios as a function of $\Sigma_\mathrm{SFR}$ (Figure~\ref{fig:sigma_sfr_stacks}), where \HCOp/\mbox{CO\Jtwo} and HNC/\mbox{CO\Jtwo} vary by $\sim0.5\,\mathrm{dex}$. The average line ratios in this case are $0.042\pm0.002$ for \HCOp/\mbox{CO\Jtwo} and $0.015\pm0.002$ for HNC/\mbox{CO\Jtwo}.

Here we list our results and compare them with previous studies of NGC~3627 at lower spatial resolution. We find a peak HCN/CO\Jtwo{} ratio of 0.076 in the very centre of NGC~3627, and a mean value of $0.058\pm0.002$ across the central 1~kpc (see Figure~\ref{fig:radial_profiles}). Similarly, across the central 1~kpc of NGC~3627, the studies of \cite{gallagher_2018b} and \cite{jimenez19}, who conduct a comparable stacking analysis, report mean a HCN/CO\Jone{} ratio of ${\sim}0.04$. The \HCOp/CO\Jtwo{} in the centre of NGC~3627 is $0.078$, and $0.047\pm0.002$ the mean value in the inner 1~kpc region. For the \HCOp/CO\Jone{} line ratio, \cite{jimenez19} found a value of 0.017 at 1~kpc from the centre, whilst \cite{gallagher_2018a} reported a value of 0.039 in the centre and 0.01 at 1~kpc from the centre. Overall, our values towards the centre agree well with previous studies at lower spatial resolution, if we take into account the mean CO\Jtwo/CO\Jone{} ratio of 0.59 in NGC~3627 \citep{denbrok_2021}. Comparing ratios determined here to the literature at larger Galactic radii is not possible, as our mosaic is not fully complete for bins $>$2~kpc (see lower right panel of Figure~\ref{fig:radial_profiles}).

\subsubsection{Line ratios of HCN, HNC and \HCOp}

In this section, we investigate the line ratios between the tracers of denser molecular gas, i.e. HNC/HCN, \HCOp{}/HCN and HNC/\HCOp{}, as a function of three different stacking properties (bottom left panels of Figures~\ref{fig:radial_profiles}, \ref{fig:co_stacks} and~\ref{fig:sigma_sfr_stacks}).

The HNC/HCN ratio is considerably lower than unity as a function of all the stacking properties. When considering only the significant points for the HNC/HCN ratio as a function of galactocentric radius (the bottom left panel in Figure~\ref{fig:radial_profiles}), the average HNC/HCN line ratio across NGC~3627 is $0.30\pm0.03$. The mean value of HNC/HCN ratio in the inner 1.2\,kpc region is $0.30\pm0.05$ which is in agreement with the one reported by \cite{jimenez19} for the inner 1.2\,kpc region of five barred galaxies ($0.4\pm0.1$). \cite{watanabe_2019} found a value of $0.3\pm0.1$. HNC/HCN does not vary with the CO\Jtwo{} emission, but we report a slightly higher value of this line ratio for CO\Jtwo{} integrated intensities below $200\,{\rm K}\kms$. The mean HNC/HCN line ratio within all the CO\Jtwo{} bins is $0.387\pm0.024$, whereas it is $0.46\pm0.06$ as a function of the $\Sigma_\mathrm{SFR}$.

As a function of the environment in NGC~3627, we find HCN/\HCOp{} greater than unity in the central region. We measure the highest value ($1.9\pm0.4$) in the bar around 2\,kpc, whereas we report values lower than unity in the bar ends where we measure the minimum value of this line ratio ($0.42\pm0.08$). At the very centre of NGC~3627 we find HCN/\HCOp{} to be $1.27\pm0.02$. Overall, we find the mean HCN/\HCOp{} to be ($1.31\pm0.06$) in the inner 1.2\,kpc region, and both \cite{jimenez19} and \cite{watanabe_2019} reported $1.3\pm0.2$. For the CO\Jtwo{} stacks (the bottom left panel in Figure~\ref{fig:co_stacks}), we find values lower than unity at CO\Jtwo{} integrated intensities below $200\,K\kms$, whereas regions with brighter CO\Jtwo{} emission have HCN/\HCOp{} values above unity. The variation of HCN/\HCOp{} is $0.5\,\mathrm{dex}$ across 2~orders of magnitude in CO\Jtwo. From the bottom left panel in Figure~\ref{fig:sigma_sfr_stacks}, we find an average HCN/\HCOp ratio of $0.91\pm0.08$. Here HCN/\HCOp varies over $0.3\,\mathrm{dex}$.

The HNC/\HCOp{} ratio has a value below unity in all three cases. The mean HNC/\HCOp{} ratio across NGC~3627 is $0.32\pm0.04$. The HNC/\HCOp{} ratio varies over $0.3\,\mathrm{dex}$ as a function of CO\Jtwo{} and over $0.2\,\mathrm{dex}$ as a function of $\Sigma_\mathrm{SFR}$.

\section{Discussion}{\label{sec:discussion}}

In this work, we observe high critical density molecules (HCN, HNC and \HCOp{}) and CO isotopologues (\thCO{} and \CeiO) across the disc of the star-forming galaxy NGC~3627 at scales of $100$~pc comparable to individual GMCs. We use CO\Jtwo{} data from PHANGS-ALMA as a bulk molecular gas tracer \citep{leroy_2021b} and extinction-corrected H$\alpha$ emission from PHANGS-MUSE as a star formation tracer (E.~Emsellem et al.\ in prep). 

The observed molecular lines show emission about $10{-}100$ times fainter than the CO\Jtwo{} line. We directly detect HCN and \HCOp{} emission in the brightest regions of NGC~3627 -- the centre and bar ends. To recover the faint emission from the observed lines and increase the signal/$\allowbreak$noise ratio, we make use of the stacking technique described in Section~\ref{sec:stacking_section}. Data are stacked by three different parameters that are measured at high resolution: galactocentric radius, CO\Jtwo{} integrated intensity, and $\Sigma_\mathrm{SFR}$. Our key results are presented in Section~\ref{sec:results}.

\subsection{Integrated intensities}{\label{sec:disc_ii}}

From the top left panel of Figure~\ref{fig:radial_profiles}, we note that there is a lack of emission along the inner part of the bar ($1.2{-}2.2$\,kpc), except for the brightest observed line in our sample, \thCO. We investigate line intensity profiles across different environments where we were able to recover emission over significantly extended continuous areas (the brightest regions in NGC~3627: the inner 1.2\,kpc region and the bar ends). In this section, we discuss how the line ratios vary within these two environments, as well as how the environment sets the star formation and denser gas fraction, and mention some caveats of our work.

Overall, we recover significant emission for all of our lines in bins of CO\Jtwo{} intensity and $\mathrm{\Sigma_{SFR}}$. We find that all lines show a positive correlation with CO\Jtwo{} emission. They follow the structure of the CO\Jtwo{} emission, and show the brightest emission in the centre and the bar ends, as well where $\mathrm{\Sigma_{SFR}}$ is enhanced. Since CO\Jtwo{} emission traces the molecular cloud scale surface density, our result is expected from the assumption that the denser molecular gas is found in regions where more gas is present.

\subsection{Dense gas fraction across different environments: the centre and the bar end}

Ratios involving HCN/$\allowbreak$\mbox{CO\Jtwo}, HNC/$\allowbreak$\mbox{CO\Jtwo} and \HCOp/$\allowbreak$\mbox{CO\Jtwo} can indicate how the gas is distributed across a range of densities, as these tracers have a high contrast of critical density (see Table~\ref{tab:line_prop} and \citealp{shirley_2015}).

The HCN/CO line ratio, which is thought to be a good indicator of the dense gas fraction $f_\mathrm{dense}$, appears to be bigger in regions of galaxies with high stellar mass surface density, high gas surface density and high dynamical pressure \citep{usero_2015, bigiel_2016, jimenez_2017b, jimenez19}. This trend has been found with observations at ${\sim}1$\,kpc resolution, whereas at the 100\,pc spaces studied here we expect to see more variations with local environment and stochasticity (due to time evolution) \citep{schruba_2010, querejeta_2019}. To first order, $f_\mathrm{dense}$ and $SFE_\mathrm{dense}$ are a function of galactocentric radius. A similar result has also been presented in recent studies at sub-kpc (${\sim}500$~pc) scales by \citet{gallagher_2018a, gallagher_2018b} and \citet{querejeta_2019}. 

\cite{gallagher_2018b} found that the HCN/$\allowbreak$CO\Jtwo{} ratio is higher in the inner kiloparsec of four nearby galaxies (barred galaxies NGC~3351, NGC~3627, NGC~4321 and unbarred galaxy NGC~4254). We find that the HCN/$\allowbreak$CO\Jtwo{} ratio is elevated in the centre of NGC~3627 by approximately a factor~2, compared to the bar end regions (see Figure~\ref{fig:radial_profiles}). One possible explanation for this elevated HCN/CO\Jtwo{} ratio, and hence higher denser gas fraction, is that the bar is driving a significant amount of gas towards the centre, which leads to an accumulation of (denser) molecular gas \citep{sheth_2005, krumholz_2015, sormani_2015a, sormani_2015b, tress_2020}. It is worth mentioning that other effects may play a role in enhancing the HCN/CO\Jtwo{} ratio (e.g. see \citealp{barnes2020}). For example, HCN emission can be enhanced relative to CO emission due to the presence of a very strong infrared (IR) emitting source (i.e. via radiative IR pumping; e.g. \citealp{matsushita_2015}). CO\Jtwo{} could have a high optical depth in some regions, which would cause a relative decrease in its emission relative to HCN emission. Moreover, the centre of NGC~3627 hosts an AGN \citep{filho_2000}.

Our results indicate that the HCN/$\allowbreak$CO\Jtwo{} ratio, interpreted to trace the dense gas fraction $f_\mathrm{dense}$, appears not to correlate with H$\alpha$ emission on ${\sim}100$\,pc scales in NGC~3627 (see Figure~\ref{fig:sigma_sfr_stacks}). \cite{gallagher_2018a} showed that $f_\mathrm{dense}$ positively correlates with the star formation efficiency of molecular gas ($SFE_\mathrm{mol}$) even though a significant scatter is present. These authors also found that $f_\mathrm{dense}$ correlates more strongly with the dense gas star formation efficiency ($SFE_\mathrm{dense}$) as compared to $SFE_\mathrm{mol}$ (see figures~6 and~7 in \citealp{gallagher_2018a}). $SFE_\mathrm{dense}$ is a strong function of the environment, i.e. it appears to be reduced in regions of high molecular gas surface density and high stellar surface density \citep{shetty_2014a, shetty_2014b}. \cite{murphy_2015} found similar results for NGC~3627: they showed that $SFE_\mathrm{mol}$ anti-correlates with $f_\mathrm{dense}$ in the centre and the bar ends. This study also suggested that the dynamical state of dense gas sets this reduced star formation efficiency.

\begin{figure}
	\includegraphics[scale=0.85]{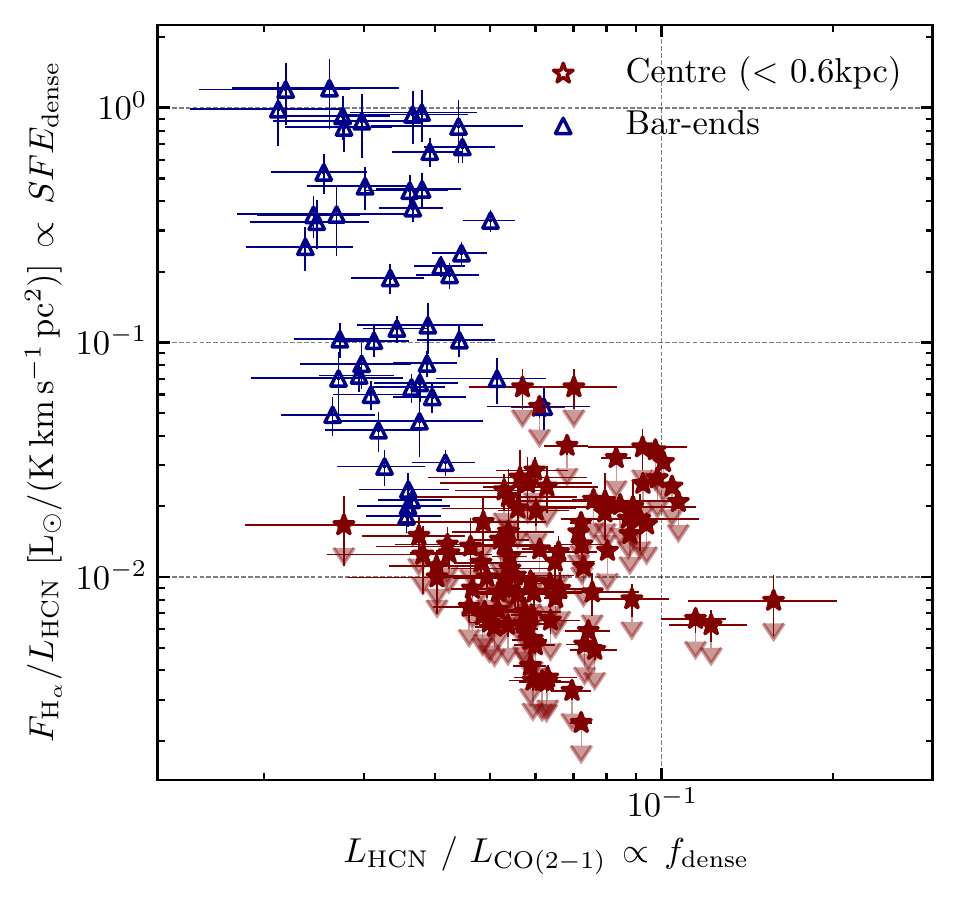}
    \caption{H$\alpha$/HCN luminosity ratio versus HCN/CO\Jtwo{} luminosity ratio which can be thought of as proxies of $SFE_\mathrm{dense}$ and $f_\mathrm{dense}$, respectively. Shown as red and blue points are independent sight lines we extract using the environmental masks \citep{querejeta_2020}. These sight lines are coming from the centre and the bar ends of NGC~3627, respectively. The central $SFE_\mathrm{dense}$ points should be considered as upper limits due to the additional contribution of H$\alpha$ emission not attributed to star formation (see Section~\ref{sec:sfr})}
    \label{fig:sfe_dense}
\end{figure}

In Figure~\ref{fig:sfe_dense}, we compare the extinction-corrected H$\alpha$/HCN luminosity ratio as a function of the HCN/CO\Jtwo{} luminosity ratio for all positions with significant CO and HCN emission. To extract sight lines from different environments, we apply environmental masks to our data: we assign points coming from the centre of NGC~3627 (inner radius of 600\,pc region) and points that belong to the spiral arm or the bar (bar ends). The points are colour-coded according to the environment from which they come. We find that central and bar end positions have relatively similar values of HCN/CO\Jtwo{} with mean $f_\mathrm{dense}\sim0.025$ and $\sim0.04$, spanning less than an order of magnitude, $f_\mathrm{dense}\sim0.015{-}0.1$. However, we find that they have significantly different H$\alpha$/HCN values with mean $SFE_\mathrm{dense}\sim0.01$ and $\sim0.1$ in the centre and bar end, respectively. We recapitulate (see Section~\ref{sec:anc_obs}), that it is, however, likely that a large fraction of the H$\alpha$ emission in the centre of NGC~3627 is not associated with star formation, rather could be attributed to AGN activity. Therefore, the H$\alpha$ emission and, hence, $SFE_\mathrm{dense}$ for the centre points, should be treated as upper limits (as highlighted in Figure~\ref{fig:sfe_dense}). We also highlight here a possibility of the obscured star formation in the centre, not traced by H$\alpha$ emission (as discussed in Section \ref{sec:anc_obs}).

The difference in these two environments may stem from the fact that the mean gas density in the centre may be much higher than the effective critical density of HCN \citep{leroy_2017a}, which increases $f_\mathrm{dense}$. However, as previously mentioned, it has been proposed that it is the \emph{relative} over-densities within the gas that are susceptible to gravitational collapse and form stars. That said, despite having a large $f_\mathrm{dense}$, the galaxy centre would have a lower $SFE_\mathrm{dense}$. Such a trend is observed within both the Milky Way's centre (e.g. \citealp{longmore_2013a, kruijssen_2014a, barnes_2017}) and other galaxy centres (e.g. \citealp{usero_2015, bigiel_2016, jimenez19}). \cite{helfer_1997} explained a high HCN/CO\Jone{} ratio in the centre of NGC~5194 by the presence of the AGN. HCN/CO\Jtwo{} ratio found in the centre of NGC~3627 could be explained by invoking suppressed star formation. We also note that at the small spatial scales (${\sim}100$\,pc) studied in this work, the scatter that can be seen in Figure~\ref{fig:sfe_dense} can be affected or driven by evolutionary effects in the gas--star cycle as the region separation lengths are ${\sim}100{-}250$~pc \citep{kruijssen_2014, kruijssen_2018, chevance_2020a}. At the centre, however, that is most likely not the case as the fragmentation length of the gas reservoir is much smaller there as shown in e.g. \cite{henshaw_2016}, unless the entire centre goes through a feeding/$\allowbreak$starburst cycle (e.g. \citealp{kruijssen_2014a, sormani_2019, barnes_2020}). Other potential mechanisms regulating and suppressing star formation in the centre could be turbulence, suggested by the observed large velocity dispersions in both CO\Jtwo{} and HCN (see Figure \ref{fig:linewidth_co21}), the effects of magnetic fields and the increased gravitational potential of disk and bulge, though their roles are less clear \citep{krumholz_2015}. With this discussion in mind, however, it should be mentioned that a difference in the chemistry of these regions could also result in the observed scatter. Galactic Centres are known to harbour higher cosmic ionisation rates than discs regions, which can lead to the formation and distinction of molecules, and increase or decrease their respective abundances (e.g. \citealp{harada_2015}). This is discussed further in the following section.

\subsection{Line ratios between HCN, HNC and \texorpdfstring{HCO$^+$}{HCO+}}

Ratios among high critical density lines, i.e. HCN, HNC and \HCOp, may encode information about density variations, radiation field, chemistry, and optical depth. The use of these line ratios to distinguish these properties is discussed in this section. 

\subsubsection{Does the HCN/HCO$^+$ ratio highlight X-ray dominated regions?}\label{sec:agn_hco_hcn}


The HCN/\HCOp{} line intensity ratio is not driven by a single process, but it is more of an interplay between radiation field, column density, and gas density \citep{privon_2015}. The HCN/\HCOp{} ratio is thought to be a good tracer of X-ray dominated regions \citep{privon_2015, murphy_2015}, because HCN abundance relative to CO and \HCOp{} can be X-ray enhanced due to the presence of AGN \citep{jackson_1993, tacconi_1994, kohno_2002, usero_2004}, even though there are some sources with an AGN that do not show this. According to \cite{meijerink_2007}, the HCN/\HCOp{} ratio can be used to discriminate between X-ray dominated (XDR) and photon dominated regions (PDR). \cite{krips_2008} found a systematic difference in HCN/\HCOp{} between AGN, and starburst dominated systems. Moreover, \cite{privon_2015} concluded that global HCN enhancement is not necessarily a tracer of an AGN, whereas the presence of AGN does enhance HCN emission.

On the one hand, both HCN and \HCOp{} are sensitive to cosmic rays. \HCOp{} is produced predominantly by the $\mathrm{CO+H_3^+}$ reaction, where $\mathrm{H_3^+}$ is created by cosmic ray ionization of $\mathrm{H_2}$ \citep{caselli_1998}. Therefore, we expect a higher \HCOp{} abundance when the cosmic ray ionization rate is high. However, very high ionization rates may boost HCN emission by increasing the number of HCN--electron collisions \citep{kauffmann_2017}.
On the other hand, \HCOp{} as an ion tends to recombine and it is sensitive to the free electron abundance, although we would still expect the cosmic ray ionization rate to boost the \HCOp. HCN abundance, however, is not dependent on the free electron abundance \citep{papadopoulos_2007}.
Besides abundance and electron density, this line ratio can also be affected by the optical depth as shown by \citet{jimenez_2017a}. This study calculated the optical depth of HCN ($\mathrm{\tau = 4.2\pm0.5}$) and \HCOp{} ($\mathrm{\tau < 5.2}$) in NGC~3627 from the HCN/H$\mathrm{^{13}}$CN and \HCOp/H$\mathrm{^{13}CO^+}$ line ratios. Moreover, if HCN and \HCOp{} are optically thick \citep{jimenez_2017a} and sub-thermally excited across the galaxy's disc \citep{knudsen_2007, Meier_2015}, then the HCN/\HCOp{} ratio may remain close to unity with small changes across the disc. 
Within resolved Galactic star-forming regions, the above effects combine to produce a spatial offset between HCN and \HCOp{} which highlights this line ratio as a tool to trace fundamentally different environmental conditions \citep[e.g.][]{pety_2017, barnes2020}. \cite{murphy_2015} reported a spatial offset between HCN and \HCOp{} peak emission in NGC~3627's bar ends, implying that excitation effects may also play a role.

There is significant evidence to suggest that NGC~3627 harbours an AGN in its centre, which is classified as LINER/$\allowbreak$Seyfert~2 \citep{ho_1997, filho_2000, cetty_2006}. NGC~3627 hosts a strong X-ray source in its centre \citep{grier_2011}, which could affect the surrounding gas and impact the HCN/\HCOp{} line ratio.

Our results show that the HCN/\HCOp{} line ratio has values greater than or close to unity in the central region of NGC~3627 (see Figure~\ref{fig:radial_profiles}). HCN/\HCOp{} ratio higher than unity is characteristic of either a high density PDR ($n_{\rm H} > 10^5\,\mathrm{cm^{-3}}$) or a low column density XDR ($N_\mathrm{H} < 3.16\times10^{22}\,\mathrm{cm}^{-2}$) \citep{meijerink_2007}. The presence of a XDR would cause CO\Jtwo/CO\Jone{} ratio being well above unity. \cite{law_2018} found a CO\Jtwo/CO\Jone{} ratio of $\sim 0.5-0.6$ in the nuclear region of NGC~3627 combining a CO\Jtwo{} observations from the Submillimeter Array (SMA) at $1\,\arcsec$ resolution and CO\Jone{} from a BIMA SONG survey \citep{regan_2001, helfer_2003} at $5\,\arcsec$ resolution. CO\Jtwo/CO\Jone{} ratio found in the centre of NGC~3627 is lower than the values found in XDR models \citep{meijerink_2007}, which is then suggestive that most of the CO\Jtwo{} emission does not come from an XDR.
Moreover, we estimate the X-ray flux from the AGN integrated over $\mathrm{2{-}10\,keV}$ \citep{ho_1997, filho_2000}. The flux is computed at a distance of approximately the beam size $\mathrm{100\,pc}$ from the AGN and it is ${\sim}2.6\times10^{-3}$ $\mathrm{erg\,cm^{-2}\,s^{-1}}$. This value is lower than the fluxes used in the \cite{meijerink_2005} models, which indicates that X-ray emission gets significantly absorbed very close ($\mathrm{<\,100\,pc}$) to the AGN, and, therefore, should not impact the \HCOp/HCN line ratio at larger radii.

We further consider possible composite XDR and PDR regions as explanations for the observed HCN/\HCOp{} ratio. \cite{privon_2015} reported HCN/\HCOp{} ratio of $1.84$ for an AGN dominated system, $1.14$ for composite (high density environments such as molecular cloud cores) and $0.88$ for starburst systems. \cite{garcia-burillo_2014} found an average HCN/\HCOp{} value of $2.5$ in nuclear centre of NGC~1068 at 35\,pc resolution. They reported the lowest HCN/\HCOp{} value of $1.3$. We report the average HCN/\HCOp{} value in the centre of NGC~3627 to be $1.31\pm0.06$ and the lowest HCN/\HCOp{} value of $1.07\pm0.07$ within the central region. According to \cite{privon_2015}, the average HCN/\HCOp{} in the central region of NGC~3627 is lying between the AGN dominated and composite systems.

X-ray sources have also been found in the bar ends of NGC~3627 \citep{wezgowiec_2012}, where we report a mean HCN/\HCOp{} ratio of $\sim0.85$. The X-ray luminosity from the bar ends integrated over $\mathrm{0{-}3\,keV}$ ($\sim3.5\times10^{38}$ and $4\times10^{38}$ $\mathrm{erg\,s^{-1}}$ for the Southern and the Northern bar ends, respectively) is comparable with the X-ray luminosity integrated over the same range from the centre ($\sim2.75\times10^{38}$ $\mathrm{erg\,s^{-1}}$) \citep{wezgowiec_2012}. The X-ray emission in the bar ends could then also be influencing the \HCOp/HCN ratio within the bar ends, but at spatial scales lower than the beam size ($\sim\,100\,\rm{pc}$).

\subsubsection{HNC/HCN ratio as a temperature tracer}\label{sec:temp_hacar}

Recently, \citet{Hacar_2020} has suggested that the HNC/HCN line ratio probes the gas kinetic temperature in the molecular ISM. This dependence was, however, determined using high spatial resolution ($<0.1$\,pc) observations of the Orion star-forming region. Hence, it is interesting to apply this probe to our NGC~3627 observations which cover a much larger dynamic range of environments both within the map and within the large beam (${\sim}100$\,pc). Theoretical studies have suggested that the observed variations in HCN and HNC emission could be chemically controlled. HCN can be produced in neutral-neutral collisions ($\mathrm{HNC + H \longrightarrow HCN + H}$), which when proceeding from left to right lowers the HNC/HCN abundance ratio. This reaction activates at a certain temperature, although there are some discrepancies between the observations and theoretical models. \cite{schilke_1992} calculated the activation temperature of 200 K, \cite{talbi_1996} found value of a 2000 K, whereas recent studies showed that this reaction could be activated at temperatures as low as $\sim20$K \citep{graninger_2014}. Gas-chemistry should also be taken into account, as it can enhance HCN abundance at temperatures between 30 and 60 K, and therefore influence HNC/HCN abundance ratio \citep{graninger_2014}.

Overall, we find the HNC/HCN line ratio to be less than unity within all three different types of bins shown in Figures~\ref{fig:radial_profiles}, \ref{fig:co_stacks}, and ~\ref{fig:sigma_sfr_stacks}. We also find that there is no correlation with CO\Jtwo{} intensity, and the HNC/HCN ratio does not vary within the star formation bins; e.g. assuming star formation could correlate with gas temperature. Moreover, we see that the HNC/HCN ratio is higher in the bar end than in the centre. If we assume the temperature dependence of HNC/HCN in our case, as derived by \cite{Hacar_2020} within the Orion integral shaped filament, we estimate a mean beam-averaged temperature in the centre and bar ends to be both 34~K. It is expected that higher temperatures should originate from the galactic centre region, e.g. due to the increased energetics (e.g. shocks) and/or radiation field within the central region. However, it is worth noting that NGC~3627 could be atypical as the bar ends are also very prominent in star formation and complex dynamics (i.e. that could also cause strong shocks; see Section~\ref{sec:dynamics}), and it is not clear that one would then expect a relative increase in temperature towards the centre, or not a chemical, but a physical (density) effect. 

Lastly, one may expect that a change in gas temperature could also affect the HCN/\HCOp{} ratio in the same way as the HNC/HCN ratio \citep{jimenez19}, albeit to a lesser degree i.e., due to the HCN abundance variation with temperature. Indeed, we do observe larger values of the HCN/\HCOp{} ratio within the centre compared to the disc. Yet, as previously discussed (Section~\ref{sec:agn_hco_hcn}), it is not clear if this is due to PDR and XDR, as opposed to the gas temperature.


\subsection{Dynamical interaction enhancing star formation in the bar ends}\label{sec:dynamics}

We now investigate dynamical effects that can affect star formation within the bar ends of NGC~3627. Bar ends are thought to be the interface of gas populating two major sets of orbits, from either the bar or the spiral arms. The bar ends happen to be the apocentres of the orbits that are elongated parallel to the bar, as the gas flows on these orbits near the bar ends \citep{athanassoula_1992}. At the apocentres, the gas slows down and piles up, which compresses the gas and enhances star formation. Orbital crowding and the intersection of gas on such orbits (e.g. via cloud--cloud collisions), followed by compression and collapse, are thought to enhance the star formation rate as well as the star formation efficiency in these regions (e.g. \citealp{benjamin_2005, lopez_2007, renaud_2015, sormani_2020}).

The role of orbital motions on star formation in the Northern and Southern bar end regions of NCG~3627 has been investigated by \cite{beuther_2017} using CO\Jtwo{} line emission as a kinematic tracer. The CO\Jtwo{} line was observed by PdBI and IRAM \mbox{30-m} \citep{paladino_2008, leroy_2009} at $1.6\arcsec$ (88\,pc) resolution. They found that the positions of the brightest emission (i.e. integrated intensity peaks) exhibit the broadest line profiles, almost all of which consist of multiple (sometimes even more than two) velocity components in the cold denser molecular gas. These components are interpreted as resulting from the particular arrangement of the gas populating both bar and spiral orbits. 

There are also a number of sight lines with lower line widths and only one velocity component. These are in agreement with the line width--size relation \citep{larson_1981, solomon_1987}, indicating the presence of distinct molecular clouds from only one set of orbits. Single-peaked spectra at the Northern bar end mainly trace the spiral component, whereas the bar component is observed in the single-peaked spectrum towards the Southern bar end \citep{beuther_2017}. They then used the evidence of multiple velocity components to argue that converging flows and the resulting gas pile-up in this scenario lead to enhanced star formation in the bar ends of NGC~3627. This may stem from the influence of the interaction with the galaxy's companion NGC~3628 \citep{chemin_2003} in the North, which might cause gas to be more strongly arranged by the spiral than by the bar. \cite{law_2018} reported higher CO\Jtwo/CO\Jone\ line ratios in the regions where the interaction is closer. The difference in magnetic field strength in bar ends may also play a role: \cite{soida_2001} found two magnetic field components in NGC~3627, with the magnetic field connected to the CO emission along the Western arm.

In this work, we confirm that the Northern and Southern bar ends of NGC~3627 contain not only bright CO\Jtwo{} but also HCN emission, implying that they are rich in both cold molecular and denser molecular gas. The Southern bar end appears to be brighter in both lines and more extended. Something similar is seen in H$\alpha$ emission suggesting that star formation is higher in this region. This supports the idea that these regions have ample fuel for intense star formation.  

\begin{figure*}
	\includegraphics[scale = 0.7]{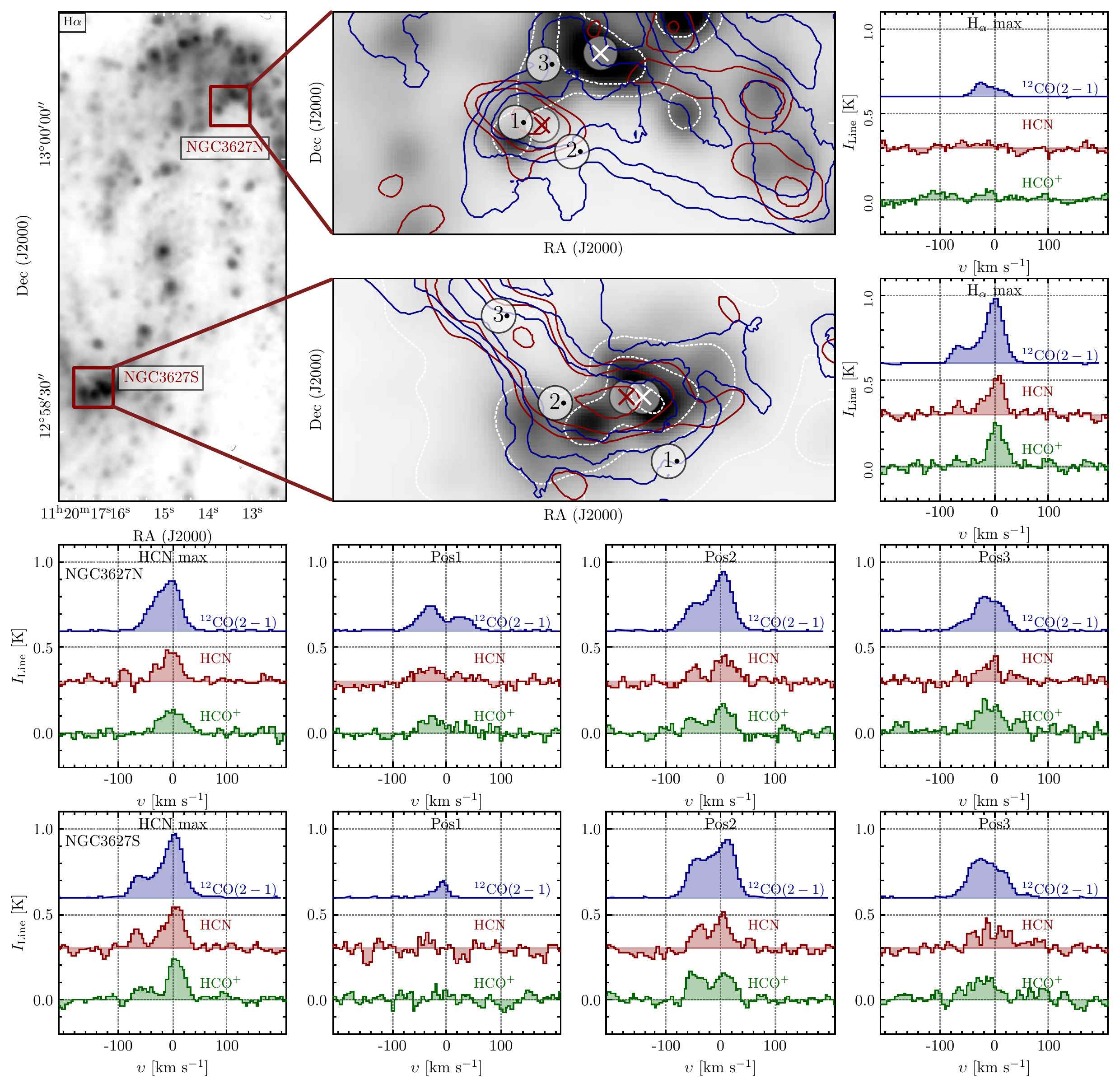}
    \caption{Bar end regions within NGC~3627. The Northern bar end (NGC~3627N) is shown in the top middle panel and the Southern bar end (NGC~3627S) in the middle panel in the second row. We show H$\alpha$ emission within these regions in grey scale and as white contours. Contours of CO\Jtwo{} emission (of 50, 100 and 200 sigma) are shown in blue, and the HCN contours of 3, 5 and 10 sigma are in red. We indicate the position of the maximum of the H$\alpha$ and HCN emission with a white cross and red cross, respectively. We show averaged spectra of CO\Jtwo, HCN and \HCOp{}. In the first two rows in right panels we plot spectra averaged over $3\arcsec$ (164~pc) region centred on the white crosses. The last two rows show averaged spectra over $3\arcsec$ (164~pc) region centered on the red crosses (left panels), while the remaining panels show averaged spectra over $3\arcsec$ regions in three positions marked with numbers 1, 2 and~3 from both bar ends (each row respectively for NGC~3627N and NGC~3627S). These positions coincide with those analysed in \citet{beuther_2017}.} 
    \label{fig:barends_spectra}
\end{figure*}

Our view of the kinematics of the denser gas in relation to the lower density gas traced by CO\Jtwo{} adds another layer to the picture proposed by \cite{beuther_2017}. Figure~\ref{fig:barends_spectra} shows a comparison between the spatial and spectral distributions of the CO\Jtwo, HCN and \HCOp{} emission at the bar ends in NGC~3627. There we also highlight the distribution of H$\alpha$ emission from MUSE observations, to serve as a probe of star formation.

The left panels in the first two rows show the H$\alpha$ maps of both bar end regions overlaid with contours of H$\alpha$, CO\Jtwo{} and HCN in white, blue and red, respectively. We find that HCN peaks in a different position than H$\alpha$ in the Northern bar end, whereas a spatial offset between these two peaks is still present but smaller in the Southern bar end.  

We take averaged spectra within regions of a single beam (${\sim}3\arcsec = 164$\,pc in radius) at several positions across both bar ends, which are presented in Figure~\ref{fig:barends_spectra}. We mark the peaks of HCN emission with red crosses and the peaks of H$\alpha$ emission as white crosses. The spectra towards these two positions in both bar ends are shown in the right panels of the top two rows and in the left panels of the bottom two rows in Figure~\ref{fig:barends_spectra}. 

The remaining panels in the bottom two rows show three different positions taken from \citet{beuther_2017} (see their Figure~3). They identified one, two and three CO\Jtwo{} velocity components in the Pos1, Pos2 and Pos3 spectra, respectively. We see two or more velocity components in CO\Jtwo{} within almost all regions shown. We also find that both the HCN and \HCOp{} lines have two velocity components within several sight lines at the same velocities as CO\Jtwo.

We find that there are multiple velocity components in both HCN and \HCOp{} emission, which appear to peak at a similar velocity and have comparable line widths to the components previously identified in CO\Jtwo{} \citep{beuther_2017}. Moreover, we find double-peaked line profiles towards the peak of the HCN integrated intensity in both bar ends, and a double-peaked profile towards the H$\alpha$ peak in the Southern bar end. This result has two important implications. The first is the confirmation of these CO\Jtwo{} components as real physical structures, rather than optical depth effects, as it is unlikely that both HCN and HCO$^{+}$ are severely optically thick and suffer from self-absorption on ${\sim}100$\,pc scales. The second is then that the velocity components observed in CO\Jtwo{}, a molecular gas tracer, have significant dense gas tracer emission. This then highlights that the region contains both molecular gas and dense molecular gas that are distinct in velocity (and presumably space). 

We ask then the question: how did these dense molecular gas components form? The profiles shown in Figure~\ref{fig:barends_spectra} could suggest two scenarios. First, the dense gas could be created at the orbit intersection. Indeed, the peak of the dense gas emission within the centre of the bar end regions would certainly point to this enhancement. Second, converging gas flows into the bar end region could already be rich with dense gas before arrival. This could be evidence by the spectrum seen at Pos1 in the southern bar, which contains both HCN and \HCOp{} emission, yet sits further upstream from the bars rotation, and the bright star-forming region. Answering this question is, therefore, somewhat speculative, and in reality, the scenario is likely that both some dense gas is delivered to the bar, which then is further compressed. 

There is a final interesting point to note within the Southern bar end region. This is that the good correlation between H$\alpha$ and HCN and CO emission at the leading edge of the bar end (i.e. centre right of the zoom-in panel in Figure~\ref{fig:barends_spectra}), yet the downstream material appears to be under-luminous in H$\alpha$ for a similar amount of HCN emission (i.e. upper left of zoom-in). This then may highlight that star formation is more efficient for a given amount of dense gas within the leading edge of the bar, compared to the trailing material. This then opens a question, if not just the amount of dense gas is important in fuelling star formation, but also the dynamics of the dense gas in limiting and/or driving its collapse? In particular, if the compressing motions of the leading bar edge are enhancing star formation, whilst the shearing forces of the trailing bar end are inhibiting star formation.

\subsection{Systematic environmental density variations}

\cite{leroy_2017a} investigated how the line intensity ratios of molecules trace molecular gas density. In extragalactic observations, reaching the necessary resolution and sensitivity while observing molecular lines fainter than CO is challenging. The line intensity integrated within a single beam encompasses a lot of different physical gas and contains information from a range of densities. The reason is that the line emits most at the `effective critical density' ($n_\mathrm{eff, crit}$) as described in \cite{leroy_2017a}, but it also emits from densities below $n_\mathrm{eff, crit}$ which frequently dominate the (sub-beam) density distribution. Therefore, line intensity ratios can be used to infer density variations. We also note that optical depth, temperature, IR pumping, cosmic rays, and element abundance variations can affect the line ratios.
\cite{leroy_2017a} coupled simple radiative transfer models and parametrized density PDFs to quantify how changes in the sub-beam density distributions affect the beam-averaged line emissivity. This study implemented two commonly considered density distributions to describe the density distribution of the cold phase of the ISM where our observed lines are produced: a pure log-normal distribution and a log-normal distribution with a power-law tail at high densities \citep{federrath_2013, girichidis_2014, kainulainen_2014, lombardi_2014, schneider_2015a, schneider_2015b, schneider_2016}. They found that the sub-beam density distribution affects the beam-averaged emissivity. This result is interpreted in a way that lines that trace denser gas (i.e. lines with high effective critical density) are the most sensitive to density changes. Therefore, ratios of these line's intensities to CO lead to more variation than lines that trace pure molecular or bulk molecular gas (\thCO, CO\Jtwo). Such a line ratio shows a different behaviour that depends on the density distribution for a fixed abundance. The shape of the distribution of molecular gas also sets the variation of line ratios. For example, HCN/CO shows greater scatter when molecular gas is described with the log-normal distribution as opposed to the case when the log-normal distribution with a power-law tail describes molecular gas.

\begin{figure}
	\includegraphics[scale=0.65]{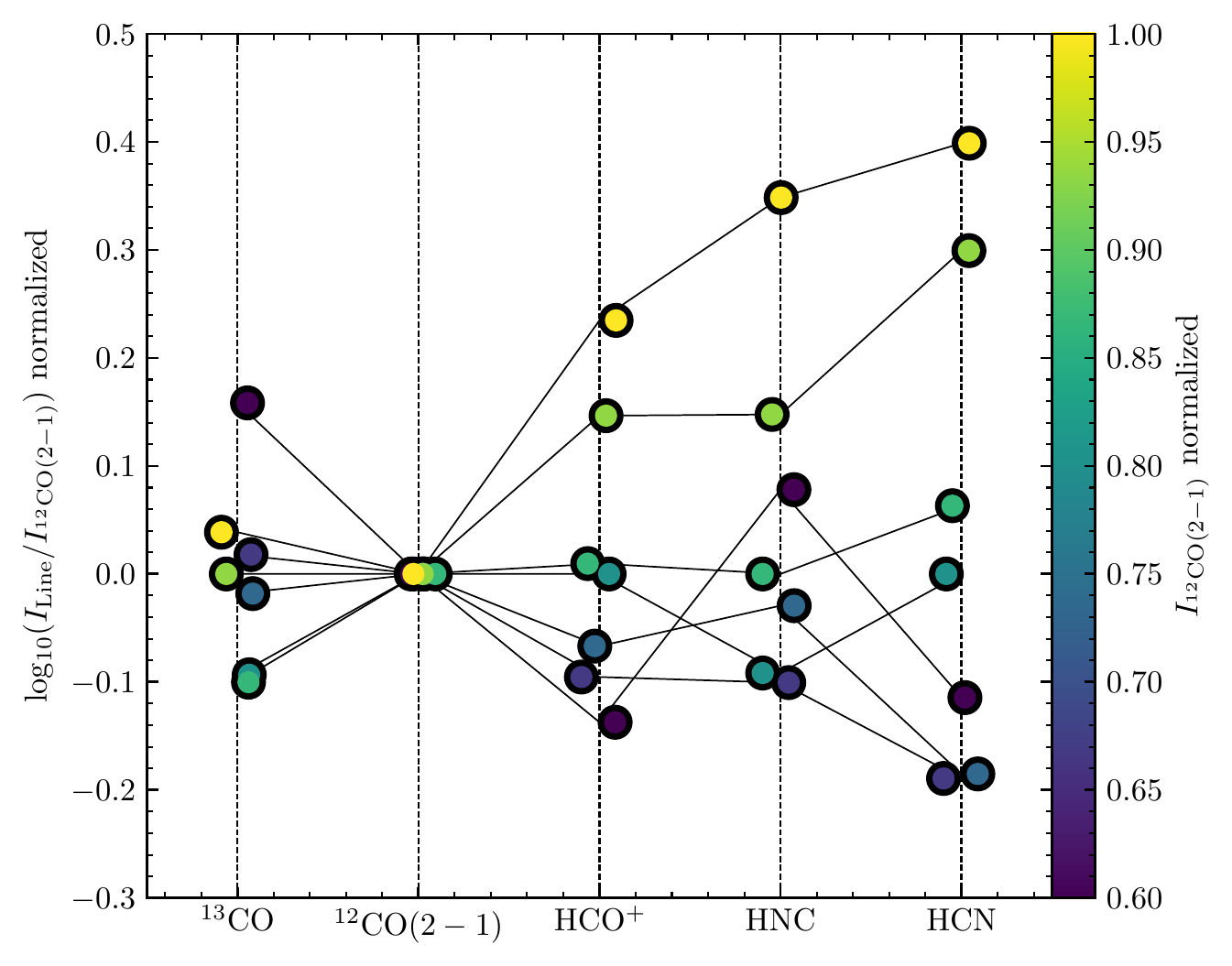}
    \caption{Line-to-CO\Jtwo{} integrated intensity ratios. Individual data points are from stacking spectra by CO\Jtwo. We only show points at which all stacked intensities have signal/$\allowbreak$noise greater than 7 (see solid points in the top right panel of Figure~\ref{fig:co_stacks}). All points for each line ratio (same x-axis label) are normalised by their respective median value. Points are colour-coded according to CO\Jtwo{} integrated intensity (normalized) and are slightly offset along the x-axis for visualisation purposes. All points with the same normalized CO\Jtwo{} value are connected with a black line.}
    \label{fig:scatters_colorcode_co21}
\end{figure}

In this section, we investigate the scenario from \citet{leroy_2017a} that line intensity ratios can reflect changes in density distributions. In Figure~\ref{fig:scatters_colorcode_co21}, we show how the line intensity ratios vary as a function of the CO\Jtwo{} integrated intensity, used as a proxy for the volume density at cloud scales \citep{leroy_2016, sun_2018}. All intensities are determined by stacking in bins of CO\Jtwo{} integrated intensity and only keeping significant measurements. We colour-code each stacked line ratio measurement by the (normalized) CO\Jtwo{} integrated intensity. Points with the same colour (similar ``density bin'') are connected with a black line (see caption for details). We sort the line-to-CO\Jtwo{} ratios according to their effective critical density listed in Table~\ref{tab:line_prop}. 

We show points at which the CO\Jtwo{} integrated intensity is above $\sim$50 K km $\mathrm{s^{-1}}$, because for these CO\Jtwo{} intensity values the stacked line integrated intensities (except for \CeiO) are classified as significant (the signal/$\allowbreak$noise is $>$ 7) according to our criteria explained in Section \ref{sec:anc_obs} (also see Figure\,\ref{fig:co_stacks}).

Overall, we find that HCN/CO\Jtwo{} shows the highest scatter, followed by HNC/CO\Jtwo{}, and \HCOp/CO\Jtwo. The flaring pattern seen in Figure~\ref{fig:scatters_colorcode_co21} is consistent with the sensitivity of these line ratios to changes in mean gas density (here as traced by cloud-scale surface density). The higher the contrast of line critical densities, the more pronounced the flaring and dynamical range. This agrees with the modelling results found in \citet{leroy_2017a} and J. Puschnig et al. (in prep).  We note that the \thCO/CO\Jtwo{} ratio shows a reversed pattern, reflecting the lower \thCO{} critical density compared to CO\Jtwo{}.


\section{Summary}{\label{sec:summary}}

We present new NOrthern Extended Millimeter Array (NOEMA) observations towards a nearby, actively star-forming disc galaxy NGC~3627. These represent the current highest resolution ($1.8\arcsec \approx 100$~pc) of molecular lines that trace denser gas across a nearby spiral galaxy. Through this work, we investigate how the denser gas, using the following tracers HCN\Jone, HNC\Jone, \HCOp\Jone, \thCO\Jone{} and \CeiO\Jone{}, is spatially distributed across NGC~3627 in comparison with the spatial distribution of bulk molecular gas (CO\Jtwo{} emission) and star formation (H$\alpha$ emission). We also investigate line profiles of denser molecular gas towards the bar ends.
We list our key results here:

\begin{itemize}
    \item We first investigate how the observed line intensities vary as a function of the CO\Jtwo{} emission and, therefore, as a function of the bulk molecular gas. We find that all lines show brighter emission towards regions of brighter CO emission. To inspect the second order deviations, we investigate the ratio of the lines with respect to CO. We find that the HCN, HNC and \HCOp{} to CO\Jtwo{} ratios positively correlate with CO\Jtwo{} emission, while the remaining lines (\thCO{} and \CeiO) show the opposite trend and do not correlate with CO\Jtwo{} emission. Moreover, we test if these line ratios can trace different density distributions and investigate the scatter in the ratio with increasing CO intensity. We find that HCN, HNC and \HCOp{} to CO\Jtwo{} line ratios show greater scatter, which suggests that they trace densities above the mean molecular gas density, but it can also be driven by the other effects.
    
    \item We investigate how the line intensities vary as a function of H$\alpha$ emission. We find that all line intensities increase towards brighter H$\alpha$ emission. Furthermore, we see that the line ratios to CO\Jtwo{} do not vary significantly with H$\alpha$ emission. Therefore, the line ratios show no correlation with recent star formation. 
    
    \item We detect significant emission in HCN, \HCOp, HNC, \thCO{} and \CeiO{} towards the brightest regions in NGC~3627 -- the centre and the bar ends. Therefore, we check for trends with morphological features that describe the galaxy. The structure of the galaxy is influencing line intensities. We find that all lines show the brightest emission towards the centre and bar ends within NGC~3627. The line ratios reflect the environment: line to CO\Jtwo{} ratios appear to be slightly higher in the centre than in the bar ends. This is a direct cause of what has been described in the former two points, i.e. looking at line intensities as a function of CO\Jtwo{} and H$\alpha$. Stacking by these properties involves multiple different environments within the bins and therefore affects the trends we have seen.
    
    \item We measure the velocity dispersion of HCN and CO\Jtwo{} line emission for sight lines towards the centre and bar ends. Overall, HCN and CO\Jtwo{} velocity dispersion increase with CO\Jtwo{} line brightness, spanning ranges from 15 to 75 \kms{} (20$-$80 \kms) in HCN (CO\Jtwo), while CO\Jtwo{} changes by a factor of 1.4. The HCN velocity dispersion is lower by a few \kms{} than the velocity dispersion measured in CO\Jtwo. Sight lines towards the centre and bar ends are located in a different part of Figure\,\ref{fig:linewidth_co21}. Sight lines towards the bar end exhibit considerably lower HCN (from 15 to 30 \kms) and CO\Jtwo{} velocity dispersions (from 20 to 35 \kms) than the sight lines from the centre (from $\sim$60 to 80 \kms{} in both HCN and CO\Jtwo), despite having similar CO\Jtwo{} intensity ($\sim700$ and 1000 K \kms{} towards bar ends and centre respectively). The difference between measured HCN and CO\Jtwo{} velocity dispersions in the centre and bar ends could be explained due to several effects, such as infrared pumping of HCN in the centre, HCN and CO\Jtwo{} populating different orbits, and the difference in the gas mean densities from the centre and bar ends.
    

    \item We have probed the role of the environment on setting the star formation and thus checked the variation of the star formation efficiency with respect to the dense gas fraction on ${\sim}100$\,pc scales across two different environments (the centre and the bar ends). We see that they have different properties, e.g. the fraction of the denser gas to the bulk molecular gas does not change in the bar end as much as in the centre. The H$\alpha$/HCN ratio, however, depends on the environment and it appears to be higher in the bar ends than in the centre. These results agree well with previous studies of $f_\mathrm{dense}$ and $SFE_\mathrm{dense}$ at coarser resolutions \citep{usero_2015, bigiel_2016, jimenez19}. One possible explanation is that HCN does not trace the same amount of the denser gas in these environments, i.e. that it belongs to different parts of the density distributions that describe molecular gas within these regions (log-normal and power-law tail in the centre and in the bar end, respectively). Another interpretation of this result is that gas dynamics set the star formation on these scales and therefore set the star formation efficiency of the denser gas. Since we only use extinction corrected H$\alpha$ emission as the star formation tracer, our results might be biased in the centre of NGC~3627.
    
    \item The dynamical effects in the bar ends of NGC~3627 can enhance collisions that trigger local star formation. To investigate this, we compare the spectra of CO\Jtwo{}, HCN and \HCOp{} towards the Northern and the Southern bar end. We find that HCN and \HCOp{} have multiple velocity components associated with the CO\Jtwo{} velocity components indicating that these gas motions coming from the spiral and the bar seen in CO\Jtwo{} \citep{beuther_2017} contain the denser gas. Furthermore, we note that denser gas can get piled up in the bar ends where it interacts and enhance the star formation.
\end{itemize}

Our work demonstrates the importance of pushing the observations towards high angular resolution and sensitivity to resolve and detect dense gas tracers on the scales of individual GMCs. In the future, we plan to further investigate the dynamics of the dense molecular gas, in order to understand the observed differences between environments in galaxies, such as the centre and the bar ends. It will be important to extend this work to even higher sensitivity observations, allowing us to begin to observe even fainter molecular lines in nearby galaxies (the high-$J$ transitions of the dense molecular gas and even fainter tracers of dense molecular gas, i.e. $\mathrm{N_2H^+}$). Doing so would further our understanding of the densest ISM phase(s), and its ability to form stars.

\section*{Acknowledgements}
This work is based on IRAM/NOEMA observations carried out under project number W17BP, and the EMPIRE large program number 206-14 with the IRAM 30m telescope. IRAM is supported by INSU/CNRS (France), MPG (Germany) and IGN (Spain). IB, ATB, FB, JP, and JdB would like to acknowledge the funding provided from the European Union's Horizon 2020 research and innovation programme (grant agreement No 726384/Empire). CE acknowledges funding from the Deutsche Forschungsgemeinschaft (DFG) Sachbeihilfe, grant number BI1546/3-1. JP and CH acknowledge support by the Programme National "Physique et Chimie du Milieu Interstellaire" (pcMI) of CNRS/INSU with INC/INP, co-funded by CEA and CNES.
The work of AKL is partially supported by the National Science Foundation under Grants No. 1615105, 1615109, and 1653300.
AU acknowledges support from the Spanish funding grants PGC2018-094671-B-I00 (MCIU/AEI/FEDER) and PID2019-108765GB-I00 (MICINN). 
ES, DL, IP, TS, and FS acknowledge funding from the European Research Council (ERC) under the European Union’s Horizon 2020 research and innovation programme (grant agreement No. 694343).
AH was supported by the Programme National Cosmology et Galaxies (PNCG) of CNRS/INSU with INP and IN2P3, co-funded by CEA and CNES, and by the Programme National “Physique et Chimie du Milieu Interstellaire” (PCMI) of CNRS/INSU with INC/INP co-funded by CEA and CNES.
CMF is supported by the National Science Foundation under Award No. 1903946 and acknowledges funding from the European Research Council (ERC) under the European Union’s Horizon 2020 research and innovation programme (grant agreement No. 694343).
K.K.\ gratefully acknowledges funding from the German Research Foundation (DFG) in the form of an Emmy Noether Research Group (grant number KR4598/2-1, PI Kreckel).
MC and JMDK gratefully acknowledge funding from the Deutsche Forschungsgemeinschaft (DFG, German Research Foundation) through an Emmy Noether Research Group (grant number KR4801/1-1) and the DFG Sachbeihilfe (grant number KR4801/2-1), and from the European Research Council (ERC) under the European Union's Horizon 2020 research and innovation programme via the ERC Starting Grant MUSTANG (grant agreement number 714907).
SG, RSK, and MCS acknowledge support from the Deutsche Forschungsgemeinschaft (DFG) via the Collaborative Research Center (SFB 881, Project-ID 138713538) ``The Milky Way System'' (sub-projects A1, B1, B2 and B8) and from the Heidelberg cluster of excellence (EXC 2181 - 390900948) ``STRUCTURES: A unifying approach to emergent phenomena in the physical world, mathematics, and complex data'', funded by the German Excellence Strategy. RSK also thanks for funding form the European Research Council in the ERC Synergy Grant ``ECOGAL -- Understanding our Galactic ecosystem: From the disk of the Milky Way to the formation sites of stars and planets'' (project ID 855130). ER acknowledges the support of the Natural Sciences and Engineering Research Council of Canada (NSERC), funding reference number RGPIN-2017-03987.
MCS acknowledges financial support from the German Research Foundation (DFG) via the collaborative research centre (SFB 881, Project-ID 138713538) ”The Milky Way System” (subprojects A1, B1, B2, and B8).
MQ acknowledges support from the research project PID2019-106027GA-C44 from the Spanish Ministerio de Ciencia e Innovaci\'on.
TGW acknowledges funding from the European Research Council (ERC) under the European Union’s Horizon 2020 research and innovation programme (grant agreement No. 694343).

This paper makes use of the following ALMA data: ADS/JAO.ALMA\#2015.1.00956.S. ALMA is a partnership of ESO (representing its member states), NSF (USA) and NINS (Japan), together with NRC (Canada), MOST and ASIAA (Taiwan), and KASI (Republic of Korea), in cooperation with the Republic of Chile. The Joint ALMA Observatory is operated by ESO, AUI/NRAO and NAOJ. The National Radio Astronomy Observatory is a facility of the National Science Foundation operated under cooperative agreement by Associated Universities, Inc.

\section*{Data availability}

The data used within this paper will be shared on reasonable request to the corresponding author.



\bibliographystyle{mnras}
\bibliography{references} 


\appendix

\section{Additional figures}

\begin{figure*}
	\includegraphics[width=\textwidth,height=0.95\textheight,keepaspectratio]{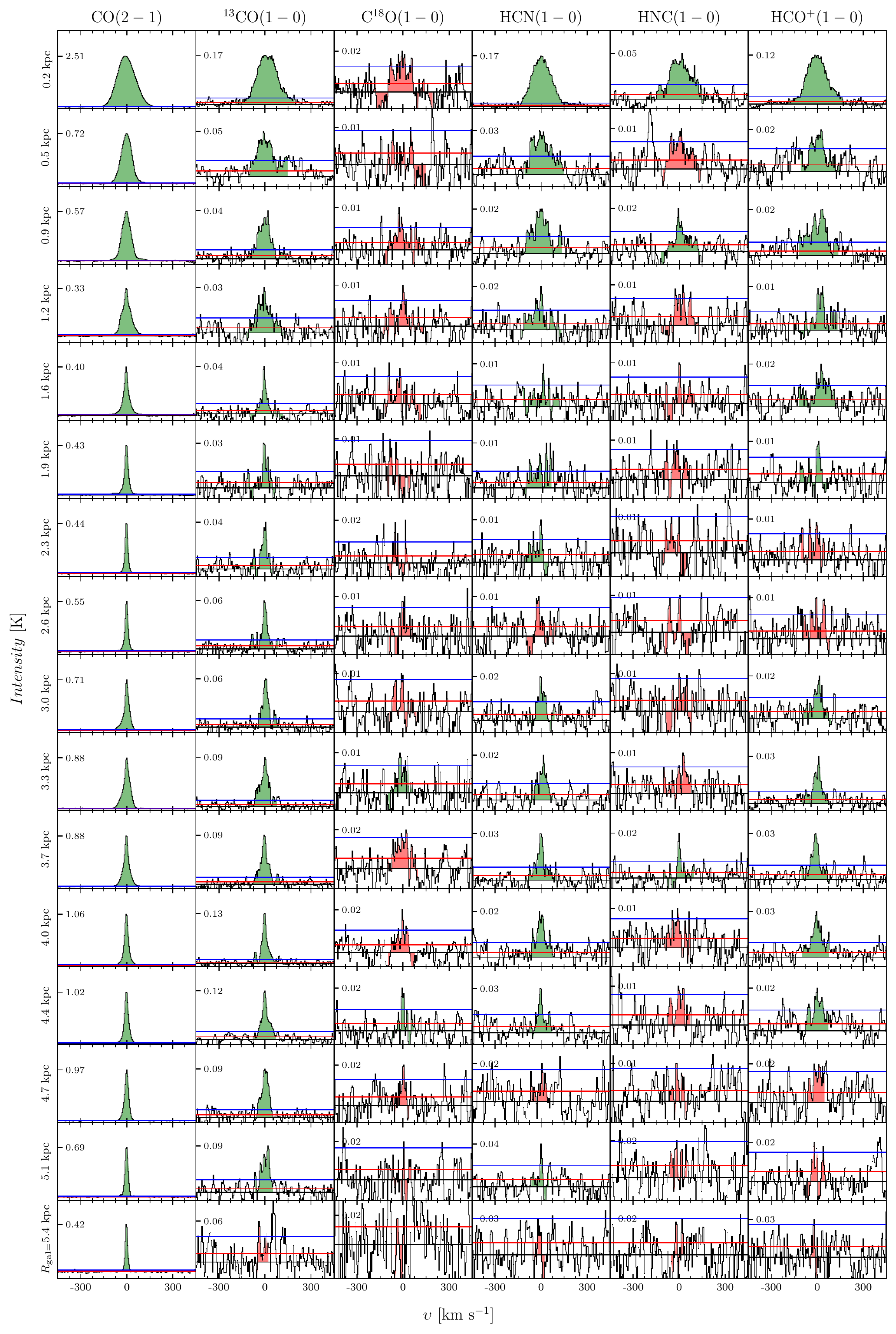}
    \caption{Radially stacked spectra of all lines (each column respectively). Median values of the radius for each bin are shown on the left side of every row. Shaded regions show the integration window defined using CO\Jtwo{} line. The colour of shaded regions correspond to whether the integrated intensity of the stacked line is defined as an upper limit (red) or not (green), as described in Section~\ref{sec:stacking_section}. The black horizontal line shows the 0-level. The red horizontal line shows the rms of the stacked profile, whereas the blue line represents 3 times the rms.}
    \label{fig:stacked_spec_rad}
\end{figure*}


\begin{figure*}
	\includegraphics[width=\textwidth,height=0.95\textheight,keepaspectratio]{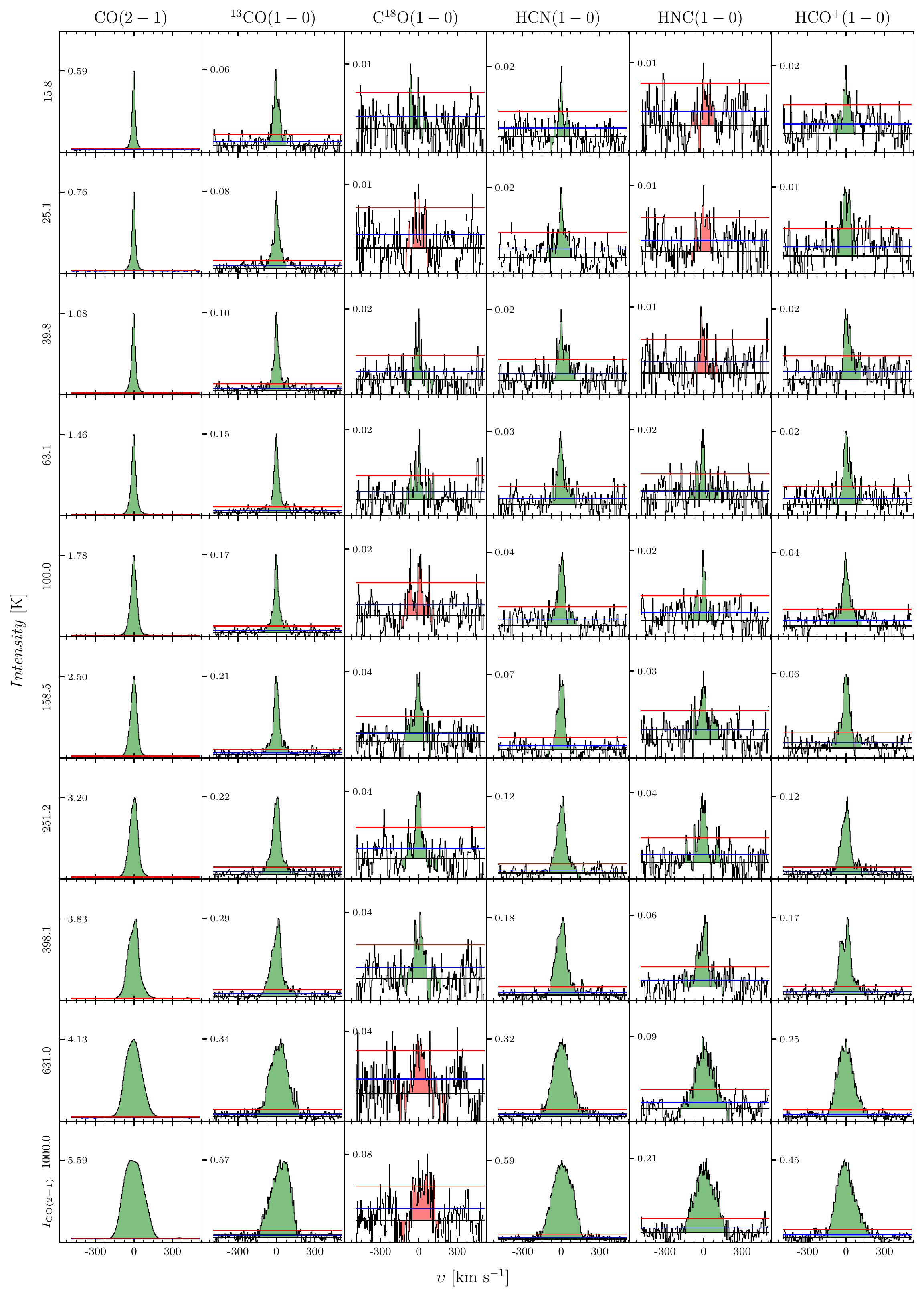}
    \caption{Same as in the Figure~\ref{fig:stacked_spec_rad}, but for stacked spectra by the $\mathrm{CO\Jtwo{}}$ integrated intensity. The CO\Jtwo{} bins are in units of K\kms.}
    \label{fig:stacked_spectra_co21}
\end{figure*}

\begin{figure*}
	\includegraphics[width=\textwidth,height=0.95\textheight,keepaspectratio]{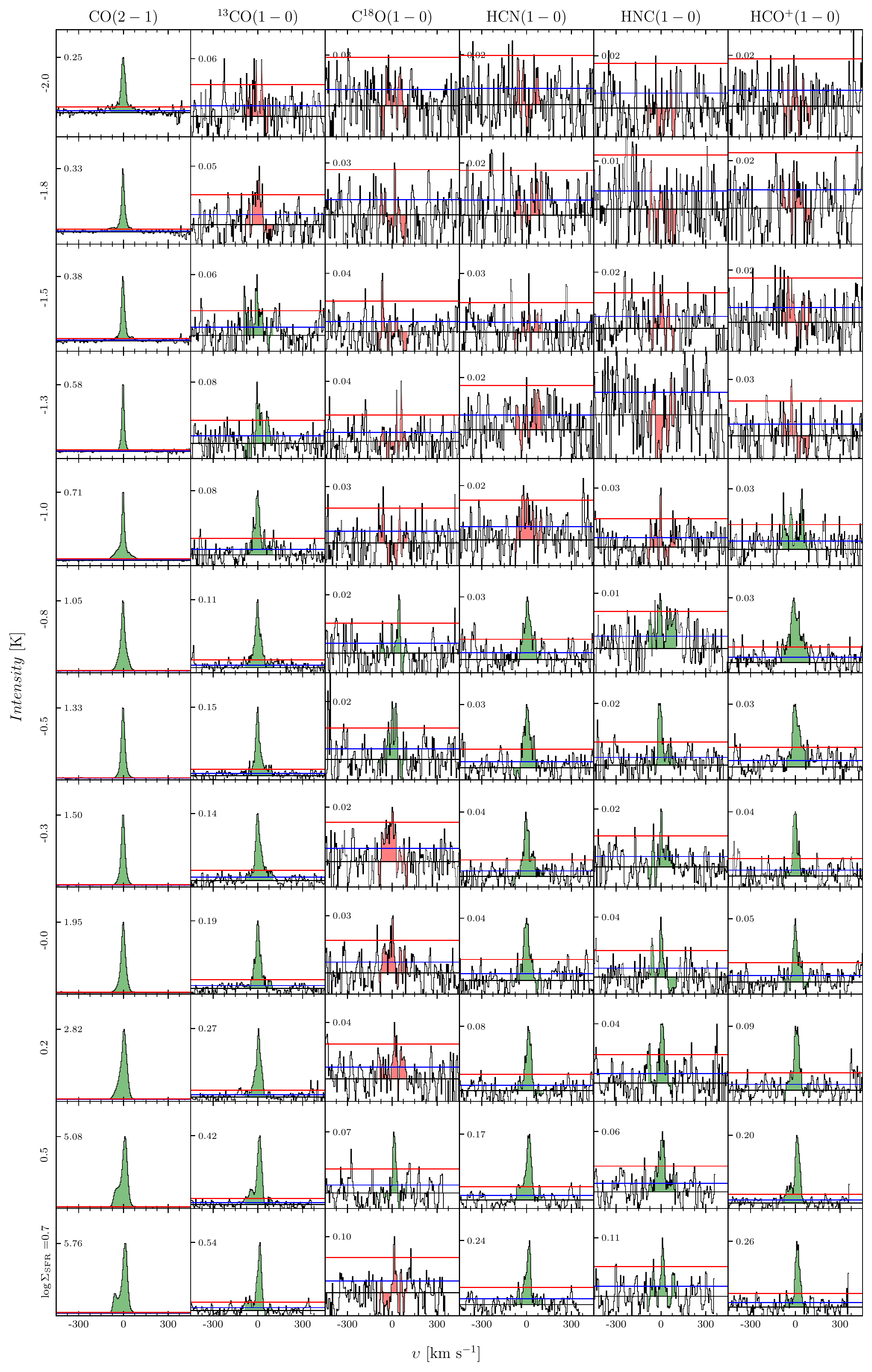}
    \caption{Same as in the Figure~\ref{fig:stacked_spec_rad}, but for stacked spectra by the $\mathrm{\Sigma_{SFR}}$. We show the $\mathrm{\Sigma_{SFR}}$ bins in logarithmic scale in units of $\mathrm{M_{\odot}\,yr^{-1}\,kpc^{-2}}$.}
    \label{fig:stacked_spectra_sfr}
\end{figure*}



\label{lastpage}
\end{document}